\documentstyle[prd,aps]{revtex}

\begin{document}
\draft

\title{The TIGA technique for detecting gravitational waves with a 
spherical antenna}

\author{Stephen M. Merkowitz\thanks{Present affiliation ROG collaboration 
and the INFN Laboratori Nazionali di Frascati, Italy} and Warren W.  
Johnson}

\address{Department of Physics and Astronomy, Louisiana State University, 
Baton Rouge, Louisiana 70803}

\date{\today}

\maketitle

\begin{abstract} 
We report the results of a theoretical and experimental study of a 
spherical gravitational wave antenna.  We show that it is possible to 
understand the data from a spherical antenna with 6 radial resonant 
transducers attached to the surface in the truncated icosahedral 
arrangement.  We find that the errors associated with small deviations from 
the ideal case are small compared to other sources of error, such as a 
finite signal-to-noise ratio.  An {\it in situ\/} measurement technique is 
developed along with a general algorithm that describes a procedure for 
determining the direction of an external force acting on the antenna, 
including the force from a gravitational wave, using a combination of the 
transducer responses.  The practicality of these techniques was verified on 
a room-temperature prototype antenna.
\end{abstract}


\section{Introduction}

Techniques to directly detect gravitational waves have been under study for 
more than 25 years~\cite{Thorne_300y}.  Two different methods are 
aggressively being pursued today: large laser interferometers, such as the 
proposed LIGO~\cite{Abramovici_Science_1992}, and cryogenic resonant mass 
antennas, such as the operating ALLEGRO~\cite{Mauceli_PRD_1996} and 
NAUTILUS~\cite{Astone_APP_1997} detectors.  While most past work on 
resonant antennas has been with the original Weber bar 
type~\cite{Weber_PR_1960}, large spherical antennas have recently been 
proposed and become of interest~\cite{OMNI_1997}.

Several characteristics of a spherical antenna make it a unique and 
interesting instrument.  First, it is omnidirectional, capable of detecting 
gravitational waves from all directions and polarizations.  In addition, 
only a single spherical antenna is necessary for determining the direction 
of an incident gravitational wave.  A sphere has a larger cross section 
than an equivalent bar \cite{Wagoner_Pavia_1976}.  A sphere can measure all 
the tensorial components of a gravitational wave, thus it is capable of 
testing different metric theories of gravity \cite{Bianchi_CQG_1996}.  
Finally, almost all of the more than 25 years of experience gained on bar 
antennas (cryogenics, resonant transducers, suspension,\ldots) can be 
applied to a spherical antenna.

We begin this paper by reviewing the interaction between an elastic sphere 
and a gravitational wave.  We start in Sec.~\ref{sec:quadrupole_gravity} by 
describing how the gravitational field can be decomposed into 5 quadrupole 
components that will have a one-to-one correspondence with the quadrupole 
modes of a sphere as described in Sec.~\ref{sec:bare_sphere}.

When we first began this problem \cite{Johnson_PRL_1993,Merkowitz_PRD_1995} 
we developed a model for a spherical antenna with 6 resonant mass motion 
sensors attached to the sphere surface at special locations.  This model 
was limited because it put relatively strong constraints on the motion 
sensors.  In order to further investigate the behavior of a more realistic 
antenna, it was necessary to generalize the model.  In 
Sec.~\ref{sec:sphere_with_res} we develop a more general description of the 
detector by keeping the number, tuning, and arrangement of the motion 
sensors arbitrary.  Section~\ref{sec:mode_channels} shows how the response 
of the motion sensors can be used to observe the 5 quadrupole modes of the 
sphere and in Sec.~\ref{sec:direction_finding} we describe how this 
information can be used to determine the direction of an external 
excitation, including that from a gravitational wave.

In Sec.~\ref{sec:ideal_TI} we show how the general equations of motion can 
be greatly simplified if a special arrangement of motion sensors is used.  
This system was the basis for our original model which we called a 
truncated icosahedral gravitational wave antenna (TIGA) 
\cite{Johnson_PRL_1993}.  The arrangement of resonators was similarly 
called the truncated icosahedral (TI) arrangement 
\cite{Merkowitz_PRD_1995}.

Other arrangements of transducers have been suggested 
\cite{Lobo_EPL_1996,Zhou_PRD_1995}, however, we feel the TI arrangement is 
advantageous not only because it simplifies the equations of motion but 
because it maintains equal sensitivity to gravitational waves from all 
directions and polarizations.  We have also found that it facilitates in 
the interpretation of the signal from the motion sensors.  In addition, it 
has been shown that in the presence of noise the use of 6 resonant 
transducers in the TI arrangement, compared to that of one for bar antenna, 
do not reduce the overall sensitivity of the antenna 
\cite{Johnson_PRL_1993,Stevenson_Amaldi_1995} and that the arrangement is 
fairly robust to the failure of a single motion sensor 
\cite{Stevenson_OMNI_1997,Stevenson_PRD_1997}.

In an actual antenna, it may be difficult to achieve a perfect TI 
arrangement.  The motion sensors can be misplaced, mistuned, etc.  To 
account for this we develop a measurement technique in 
Sec.~\ref{sec:real_TI} that takes into account any small deviations from 
the ideal arrangement.  The results of a numerical simulation are also 
presented that show this technique to be accurate within reasonable levels 
of precision set for the detector components.

We begin discussion of experiments performed on a prototype spherical 
antenna in Sec.~\ref{sec:prototype}.  The results of the prototype without 
resonators attached is reviewed in Sec.~\ref{sec:uncoupled_prototype}.  The 
behavior of the prototype with resonant transducers attached is presented 
in Sec.~\ref{sec:coupled_prototype}.  To demonstrate the validity of the 
TIGA techniques in a single test, we show that from the response of the 
motion sensors, we can determine the location of an impulse excitation 
applied to the prototype's surface.  This same procedure can be applied to 
determine the direction of a gravitational wave as discussed in 
Sec.~\ref{sec:direction_finding}.

\section{Quadrupole decomposition of the gravitational field}
\label{sec:quadrupole_gravity}

A gravitational wave is a traveling time-dependent deviation of the metric 
tensor, denoted by $h_{\mu\nu}$.  We follow a common textbook development 
for the metric deviation of a gravitational wave, which finds that only the 
spatial components, $h_{ij}$, are non-zero, and further can be taken to be 
transverse and traceless \cite{Thorne_300y}.  The tensor is simplified if 
we initially write it in the ``wave-frame'', denoted by primed coordinates 
and indices.  It is a coordinate frame with origin at the center of mass of 
the detector, and the $z'\text{-axis}$ aligned with the propagation 
direction of the wave.  Since we restrict ourselves to detectors much 
smaller than the gravitational wavelength, only the time dependence of 
$h_{i' j'}$ will have significant physical effects.  Thus, the most general 
possible form for the spatial components of the metric deviation in the 
wave-frame can be written as:
\begin{equation}
   h_{i' j'}(t) = \left[ 
   \begin{array}{ccc}
      h'_+(t)      & h'_\times(t) & 0 \\
      h'_\times(t) &  -h'_+(t)    & 0 \\
      0            & 0            & 0
   \end{array} \right]
\end{equation}
where $h'_+(t)$ and $h'_\times(t)$ are the time-dependent gravitational 
wave amplitudes for the two allowed states of linear polarization, and are 
called the plus and cross amplitudes.

The detector is more easily described in the ``lab-frame'', denoted by 
unprimed coordinates and indices, with origin also at the center of 
mass of the detector, and $z\text{-axis}$ aligned with the local 
vertical.  In this frame, the primary physical effect of a passing 
gravitational wave is to produce a time dependent ``tidal'' force 
density $f^{\text{GW}}(\bbox{x},t)$ on material at coordinate location 
$x_i$ with mass density $\rho$, which is related to the metric 
perturbation by
\begin{equation}
   f^{\text{GW}}_i(\bbox{x},t) = \frac{1}{2} \rho \sum_{j} 
   \frac{\partial^2 h_{ij}(t)}{\partial t^2} x_j.
   \label{eqn:accel_metric_pert}
\end{equation}
We notice that this force can be written as the gradient of 
a time-dependent scalar potential:
\begin{equation}
   f_i^{\text{GW}}(\bbox{x},t) = \nabla_i \Phi(\bbox{x},t) = 
   \nabla_i \left( {\sum\limits_{j,k}
   {\frac{1}{4} \rho x_j \ddot{h}_{jk}(t) x_k}} \right).
   \label{eqn:metric_pot}
\end{equation}

This scalar potential is a quadratic form in the spatial coordinates.  It 
is natural to look for an alternate expression that separates the 
coordinate dependence into radial and angular parts.  Because the tensor 
$h_{ij}$ is traceless, the angular expansion can be done completely with 
the five ordinary spherical harmonics of order 2, which we denote by 
$Y_m(\theta,\phi)$ or $Y_m$.  We call the resulting time dependent 
expansion coefficients, denoted by $h_m(t)$, the ``spherical gravitational 
amplitudes.'' They are a complete and orthogonal representation of the 
cartesian metric deviation tensor $h_{ij}(t)$.  They depend only on the two 
wave-frame amplitudes and the direction of propagation, and are defined by
\begin{equation} 
   \Phi(\bbox{x},t) = \sqrt{\frac{\pi}{15}} \rho r^2 
   \sum\limits_m{\ddot{h}_m(t) Y_m}.
\end{equation}

The spherical harmonics $Y_m$ can be any linear combination or rotation of 
the standard spherical harmonics of order 2, as long as the orthogonality 
between them is maintained.  The advantage of not using the standard 
spherical harmonics will become apparent later, but for an example of an 
alternative set and their relation to a lab frame see 
reference~\cite{Merkowitz_PRD_1995}.

The 5 orthogonal spherical amplitudes $h_m$ are a complete set of 
measurable quantities of the local gravitational field.  Once a proper 
relation to the lab coordinate system is defined, the determination of the 
source direction follows immediately by inversion of this relationship 
\cite{Nadja_RAS_1995}.

By examining Eq.~(\ref{eqn:metric_pot}) closely, we note that because the 
force is a quadratic it is the equation of an ellipsoid.  Therefore, we can 
picture a gravitational wave as a time-dependent ellipsoidal deformation of 
the local coordinates.  While this may seem obvious, we note it here as 
this is a very useful visual tool for understanding the interaction between 
a gravitational wave and the detector.  We discuss this idea in more detail 
below to show the connection with other quadratic quantities.

\section{The uncoupled sphere}
\label{sec:bare_sphere}

The mechanics of a general antenna can be described by ordinary elastic 
theory.  Forces acting on the body will cause a deformation described by 
the displacement vector $\bbox{u}(\bbox{x},t)$, where $\bbox{x}$ is the 
equilibrium position of a mass element.  The equations of motion are then
\begin{equation}
   \rho \frac{\partial ^2 \bbox{u}}{\partial t^2} = 
   (\lambda+\mu) \bbox{\nabla} (\bbox{\nabla} \cdot  \bbox{u}) + 
   \mu \bbox{\nabla} ^2 \bbox{u} + \sum \bbox{f},
   \label{eqn:general_eom}
\end{equation}
where the Lam\'{e} coefficients $\lambda$ and $\mu$ specify the 
elastic stiffness of the material and $\sum \bbox{f}$ represents  
the sum of external force densities acting on the body \cite{LL_v7}.

In this paper, we include two forces in $\sum \bbox{f}$.  First, the signal 
or gravitational force density $\bbox{f}^{\text{GW}}$ from 
Eq.~(\ref{eqn:accel_metric_pert}).  Second, if objects are attached to the 
antenna, there will exist a reaction force between the object and the 
surface of the antenna.  Thus we choose to express the coupling to other 
objects, such as secondary resonators, as if they were external forces in 
Eq.~(\ref{eqn:general_eom}).  This device lets us partition the equations 
of motion in a convenient way.

A solution to the differential Eq.~(\ref{eqn:general_eom}) can be found by 
the standard eigenfunction expansion.  This allows a separation of the 
spatial and time dependence of the displacement vector
\begin{equation}
  \bbox{u}(\bbox{x}_i,t) = \sum\limits_m {a_m\left( t \right) 
  \bbox{\Psi}_m\left({\bbox{x}_i} \right)}.
\end{equation}
Each spatial eigenfunction, $\bbox{\Psi}_m\left( {\bbox{x}} \right)$, is 
the time independent part of the solution for unforced harmonic oscillation 
at the eigenfrequency $\omega _m$, and is found by solving
\begin{equation}
  -\rho \omega_m^2\bbox{\Psi}_m\ =\ (\lambda+\mu) \bbox{\nabla} 
  (\bbox{\nabla} \cdot 
  \bbox{\Psi}_m)\ +\ \mu \nabla ^2\bbox{\Psi}_m, 
\end{equation}
subject to the time-stationary boundary conditions, which for a sphere 
require that the total force per unit area at the surface vanish in the 
direction normal to the surface.  The quantity $a_m(t)$ is the 
time-dependent mode amplitude.  The mode index, m, enumerates the discrete 
set of modes, which obey the usual orthogonality property
\begin{equation}
  \int_V{\bbox{\Psi}_m(\bbox{x}) \cdot \bbox{\Psi}_n(\bbox{x}) d^3x}
  = N_m \delta_{mn}.
\end{equation}
The normalization constant $N_m$ is arbitrary, however, in the case of a 
sphere of radius $R$ we define it to be
\begin{equation}
  N_m \equiv \frac{4}{3} \pi R^3.
   \label{eqn:normalization}
\end{equation}

Combining the equations above, and using orthogonality to eliminate 
the summation, we find the standard result, one forced harmonic 
oscillator equation for each mode amplitude,
\begin{equation}
   \ddot a_m\left(t\right)+\omega_m^2a_m\left(t\right) = 
   {1 \over {\rho N_m}}\int {\bbox{\Psi}_m(\bbox{x})\cdot 
   \sum \bbox{f}\left( {\bbox{x},t} \right)d^3x}. 
   \label{mode_amplitude}
\end{equation}
The mode amplitudes are a complete set of collective coordinates for 
the description of the antenna motion.  All the interactions with the 
outside world, including gravitation, can be included as separate 
terms in the ``effective force'' on each mode.  An efficient 
approximation scheme will use only those modes needed for an accurate 
description of the antenna.  Only a few of the ``overlap integrals'' 
with $\bbox{f}^{\text{GW}}$ in Eq.~(\ref{mode_amplitude}) are 
large, therefore, only a few of the mode amplitudes are strongly coupled 
to gravitational waves.

Let us consider the case a perfectly homogeneous and isotropic sphere 
uncoupled from the outside world.  This should give a reasonable 
approximation to the behavior of a sphere where all the external forces 
are small.  The eigenfunctions for this case were found over a hundred 
years ago \cite{Jaerisch_1880,Lamb_1882}, however, more elegant 
derivations, using modern notation are available 
\cite{Wagoner_Pavia_1976,Ashby_PRD_1975}.

The eigenfunctions of a sphere can be described in terms of spherical 
harmonics $Y_{\ell m}\left( {\theta, \phi } \right)$.  Looking at the 
overlap integral in Eq.~(\ref{mode_amplitude}), we see that we 
need only consider odd-parity modes.  For a sphere of radius $R$ the 
eigenfunctions are written:
\begin{equation}
  \bbox{\Psi}_{\ell m} = \left[ {\alpha_\ell(r) 
  \bbox{\hat r} + \beta_\ell(r) R \bbox{\nabla} } \right]
  Y_{\ell m}(\theta, \phi),\ \ell \ \text{even.}   
   \label{eqn:sphere_eigf}
\end{equation}
The radial eigenfunctions $\alpha _\ell \left( r \right)$ and $\beta _\ell 
\left( r \right)$ determine the motion in the radial and tangential 
directions respectively.  There are 5 quadrupole modes of vibration which 
strongly couple to the force density of a gravitational wave, and are all 
degenerate, having the same angular eigenfrequency $\omega _o$.  They are 
distinguished only by their angular dependence.  For the remainder of this 
discussion we only consider the quadrupole ($\ell =2$) modes so we drop the 
$\ell $ in our notation.

The radial eigenfunctions are given by Ashby and Dreitlein 
\cite{Ashby_PRD_1975}:
\begin{eqnarray}
   \alpha(r)   & = & 
   cR \frac{\partial}{\partial r} j_2(qr) + 6dR \frac{1}{r} j_2(kr) 
   \label{eqn:alpha_ef_def} \\
   \beta(r) & = & 
   c j_2(qr) + d \frac{\partial}{\partial r}
   \left[ r j_2(kr) \right],
   \label{eqn:beta_ef_def}
\end{eqnarray}
where $j_2$ is the spherical Bessel function of order 2.  The longitudinal 
and transverse wave vectors are given by $q^2 = \rho \omega_o^2/(\lambda + 
2\mu)$ and $k^2 = \rho \omega_o^2/\mu$ respectively.  The boundary 
conditions
\begin{equation}
   \left.  c\frac{d}{dr} \left[{ \frac{j_2(qr)}{r} }\right] + 
   d\left[{ \frac{5}{r^2} - \frac{k^2}{2} - \frac{1}{r}\frac{d}{dr} 
   }\right] j_2(kr) \right|_{r=R} = 0,
\end{equation}
\begin{equation}
   \left. c\left[{ \frac{6}{r^2} - \frac{k^2}{2} - \frac{2}{r}
   \frac{d}{dr} }\right] j_2(qr) + 6d\frac{d}{dr} 
   \left[{ \frac{j_2(kr)}{r} }\right] \right|_{r=R} = 0,
\end{equation}
determine the uncoupled mode frequency $\omega_o$.  Inclusion of the 
normalization condition Eq.~(\ref{eqn:normalization}) determines the 
constants $c$ and $d$.  These coefficients specify the shape of the 
eigenfunctions and are all weakly dependent on Poisson's 
ratio~\cite{Wagoner_Pavia_1976,Merkowitz_PRD_1995}.

The gravitational effective force for mode m of the sphere,
$F_m^S$, from Eq.~(\ref{mode_amplitude}) is
\begin{equation}
   F_m^S \equiv \int_{V_o} {\bbox{\Psi}_m \cdot 
   \bbox{f}^{\text{GW}}\,d^3x}.
\end{equation}
Solving the integrals, using Eqs.~(\ref{eqn:metric_pot}) and 
(\ref{eqn:sphere_eigf}), we find
\begin{eqnarray}
   F_m^S(t) 
   & = & 
   \sqrt{\frac{4\pi}{15}} \rho \ddot{h}_m(t) R^4 \left[ 
   {cj_2\left( {qR} \right)+3dj_2\left( {kR} \right)} 
   \right] \nonumber \\
   & = & 
   \frac{1}{2} \ddot{h}_m(t) \, m_S \, \chi R .
\label{eqn:eff_force}
\end{eqnarray}
Thus we find that each spherical component of the gravitational field 
determines uniquely the effective force on the corresponding mode of a 
sphere, and they are all identical in magnitude.  We can interpret the 
effective force $F^S_m$ in each mode as the product of: the physical mass 
of the sphere $m_S$, an effective length $\chi R$, and the gravitational 
acceleration $\frac{1}{2}\ddot{h}_m$.  The factor $\chi$ is a weak function 
of Poisson's Ratio~\cite{Merkowitz_PRD_1995}.

We now see why it was convenient to write the gravitational wave amplitudes 
in terms of spherical harmonics: we have a clear way to make the connection 
between the gravitational strain $h_m$, the force they apply to the sphere 
modes $F_m$, and the amplitudes of the sphere's quadrupole modes $a_m$.  
There is a one-to-one correspondence between these three quantities when 
the same set of spherical harmonics are used.  Once we know any of these 
quantities, we can immediately infer the other two.  Later in this paper we 
will add one more quantity to this list, ``mode channels'' which are 
constructed from the observables of the antenna to have a one-to-one 
correspondence with the above quantities.

As in the case of the quadratic form of the gravitational field discussed 
above, the spherical harmonics $Y_m$ can be any linear combination or 
rotation of the standard spherical harmonics of order 2, as long as 
orthogonality between them is maintained.  The advantage here is that if 
the 5 quadrupole modes are not degenerate, but have ``fixed'' themselves in 
a particular orientation, one can choose an appropriate set of spherical 
harmonics that match the actual orientation of the quadrupole modes 
relative to the lab frame.  This basis set can then be used to describe the 
gravitational field to maintain the one-to-one connection between the 
spherical amplitudes and the 5 quadrupole modes of the sphere.

Also analogous to the spherical amplitudes, the deformation of the sphere 
due to the excitation of a quadrupole normal mode can be described by the 
quadratic equation of an ellipsoidal surface.  We recall that the geometry 
of an ellipsoid can be visualized by the principal axis theorem.  It shows 
that the general ellipsoid has three orthogonal axes that pierce the 
surface at three principal radii, two of which are extremal points on the 
surface.  The orientation of the axes are described by 3 parameters, such 
as Euler angles.  The shape is described by the relative size of the 
principal radii.  If we call $dr_1$, $dr_2$, and $dr_3$ the deviation of 
these radii from their average, then a true sphere has $dr_1=dr_2=dr_3=0$, 
an oblate (or prolate) ellipsoid has $dr_2=dr_3=-2dr_1$, and a triaxial 
ellipsoid has $dr_1>dr_2>dr_3$.  Thus 6 parameters completely describe the 
geometry.  However, there is one restriction on the ellipsoids describing 
the quadrupole modes: they are isovolumetric with the sphere, which 
requires $dr_1+dr_2+dr_3=0$, so that 5 parameters suffice.  Since the 
superposition of any ellipsoid is another ellipsoid, the eigenfunctions 
$Y_{1} \ldots Y_{5}$ form a complete and orthogonal basis set for a 5 
dimensional abstract vector space that describes all possible isovolumetric 
ellipsoids and all possible quadrupolar vibrations of the sphere.

\section{Sphere coupled to an arbitrary number of resonators}
\label{sec:sphere_with_res}

We have just shown that measurement of the quadrupole modes of a sphere 
measures all of the tensorial components of the gravitational field, but a 
simple spherical resonator is not a practical detector.  One requirement 
for practicality is a set of secondary modes or mechanical resonators.  All 
current bar antennas use resonators that interact only with the vector 
component of antenna motion normal to the surface on which they are 
mounted, thus it seems natural to restrict our consideration to resonators 
of this type.

We choose here to describe the sphere's quadrupole modes in the coupled 
system using the eigenfunctions derived above for the uncoupled sphere.  
Lobo and Serrano showed this approximation to be valid when the ratio of 
the mass of the sphere to the mass of a resonator is much less than one 
\cite{Lobo_EPL_1996}.  This approximation allows us to use a much more 
simple mathematical framework, without loss of generality, as all the 
proposed detectors \cite{OMNI_1997} satisfy this requirement.

We look now at $J$ number of resonators attached to the sphere surface at 
arbitrary angular positions $\left({\theta_j,\phi_j} \right)$.  The values 
of the relative radial displacements of the sphere surface at the resonator 
locations can be grouped together into a ``pattern vector'' for a 
particular mode.  These column vectors in turn may be collected together to 
form a ``pattern matrix'' $B_{mj}$ defined by
\begin{equation}
   B_{mj} \equiv
   \frac{1}{\alpha} \hat{\bbox{r}} \cdot 
   \bbox{\Psi}_m\left({\theta_j,\phi_j}\right),
   \label{eqn:pattern_matrix_def}
\end{equation}
where $\alpha$ is the radial eigenfunction given by 
Eq.~(\ref{eqn:alpha_ef_def}) evaluated at the surface of the sphere.  From 
Eq.~(\ref{eqn:sphere_eigf}) we find
\begin{eqnarray}
   B_{mj} = Y_m\left({\theta _j,\phi _j} \right).  
\end{eqnarray}
Because the eigenfunctions are invariant to reflection through the origin, 
we may restrict the location of resonators to one hemisphere, without loss 
of generality.

By mechanical resonator we mean a small elastic system that has 
one of its own normal modes tuned to be resonant with the 
frequency of the antenna.  The antenna surface motion excites 
this mode, and there is resonant transfer of momentum between the 
resonator and the antenna.  Hence it acts as a resonant 
mechanical transformer, turning small motions of the large 
antenna into large motions of the small resonator.  Each 
resonator $j$ is constructed to obey a one-dimensional harmonic 
oscillator equation:
\begin{equation}
  m^r_j {\ddot q}_j(t) + 
  m^r_j \sum\limits_m {\alpha B_{mj} {\ddot a}_m(t)} + 
  k^r_j {q}_j(t) 
  =
  F^r_j(t).
  \label{eqn:resonator_eom}
\end{equation}
The displacement of a resonator, relative to the sphere surface, is 
denoted by $q_j$.  Any random or noise forces that act between the small 
resonator and the sphere are included in $F^r_j$.  Under ideal 
circumstances we would assume that the resonators are identical, such that 
the mass $m^r_j$ and spring constant $k^r_j$ of each are tuned to match the 
frequency of the degenerate 5 sphere modes, however, at this point we wish 
to keep the equations general so we {\it do not\/} put any restrictions on 
these parameters.

Combining the above, we find the coupled equations of motion for the 
sphere modes are
\begin{equation}
  m^s_m {\ddot a}_m(t) +
  k^s_m a_m(t) - 
  \sum\limits_j{\alpha B_{mj} k^r_j q_j(t)}
  =
  - \sum\limits_j{\alpha B_{mj} F^r_j(t)} + F^s_m(t).
  \label{eqn:coupled_eom}
\end{equation}
Again, under ideal circumstances we would assume that the 5 quadrupole 
modes of the sphere are degenerate so that the mass $m^s_m$ and spring 
constant $k^s_m$ of each mode are identical, however, at this point we wish 
to keep the equations general so we {\it do not\/} put any restrictions on 
these parameters.

It is convenient to combine Eqs.~(\ref{eqn:resonator_eom}) and 
(\ref{eqn:coupled_eom}) into a matrix notation.  We denote matrices by a 
double underscore and column vectors by a single underscore.  We begin by 
defining the following diagonal matrices:
\begin{eqnarray*}
	\begin{array}{ll}
		M^s_{jm} \equiv \delta_{jm} m^s_m, \;&
		M^r_{jm} \equiv \delta_{jm} m^r_j, \\
		K^s_{jm} \equiv \delta_{jm} k^s_m, \;& 
		K^r_{jm} \equiv \delta_{jm} k^r_j.
	\end{array}
\end{eqnarray*}
The complete set of coupled equations of motion can now be written:
\begin{equation}
	\left[{ \begin{array}{cc}
		\underline{\underline M}^s & 
		\underline{\underline 0} \\
		\alpha 
		\underline{\underline M}^r 
		\underline{\underline B}^T & 
		\underline{\underline M}^r
	\end{array} }\right]
	\left[ \begin{array}{c}
		\underline{\ddot a}(t) \\
		\underline{\ddot q}(t)
	\end{array} \right] 
	+
	\left[ \begin{array}{cc}
		\underline{\underline K}^s & 
		-\alpha \underline{\underline B} \, 
		\underline{\underline K}^r \\
		\underline{\underline 0} & 
		\underline{\underline K}^r
	\end{array} \right]
	\left[ \begin{array}{c}
		\underline a(t) \\
		\underline q(t)
	\end{array} \right]
	=
	\left[ \begin{array}{cc}
		\underline{\underline I} & 
		-\alpha 
		\underline{\underline B} \\
		\underline{\underline 0} & 
		\underline{\underline I}
	\end{array} \right]
		\left[ \begin{array}{c}
		\underline F^s(t) \\
		\underline F^r(t)
	\end{array} \right].
\label{eqn:eom_matrix}
\end{equation}
The column vector $\underline{a}$ has 5 components, one for each sphere 
mode, and the column vector $\underline{q}$ has $J$ components, one for 
each resonator.  The dimensions of the constant matrices can be inferred 
from these two column vectors.  The matrix $\underline{\underline{0}}$ is 
defined to have all elements equal to zero, and $\underline{\underline{I}}$ 
is the identity matrix.
 
These equations should give a good account of the mechanics of the system 
for arbitrary numbers and locations of resonators.  We do not include 
terms which represent the ``dissipation'' part of friction, which can be 
shown to be negligible for the calculations we do here, however, we do 
include the ``fluctuation'' part of friction, within the random driving 
forces in $\underline{F}^s$ and $\underline{F}^r$.

We also do not include any deviations to the shape of the quadrupole modes.  
One possible cause for changes in shape is the the attachment of the 
resonators.  This would obviously become a problem if very large resonators 
were used.  However, as shown by Lobo and Serrano \cite{Lobo_EPL_1996}, if 
we limit ourselves to resonators with mass less than 1\% of the sphere 
mass this effect becomes negligible.  A second possible cause for a change 
in mode shape is a hole drilled through the sphere for suspension.  
However, finite element analysis of a sphere with a hole 
\cite{Merkowitz_Thesis} as well as experiments 
\cite{Merkowitz_PRD_1996,Coccia_PLA_1996} have shown the mode shapes to be 
changed by less than 1\% due to the suspension hole.

It is clear that Eq.~(\ref{eqn:eom_matrix}) represents a set of elastically 
coupled harmonic oscillators with driving forces.  The apparent 
peculiarities (off-diagonal terms in the mass matrix and asymmetry in the 
elastic matrix) are simply artifacts of use of the non-inertial coordinates 
$\underline{q}$.  However, we can greatly simplify these equations by 
transforming to a normal coordinate system.  We begin by noting that 
Eq.~(\ref{eqn:eom_matrix}) is of the form
\begin{equation}
	\underline{\underline{M}} \, 
	\underline{\underline{\gamma}} \,
	\underline{\ddot{y}}(t) + 
	\underline{\underline{K}} \,
	\underline{\underline{\gamma}} \, 
	\underline{y}(t) 
	= 
	\underline{\underline{R}} \, 
	\underline{F}(t),
	\label{eqn:eom_form}
\end{equation}
where we have defined
\begin{eqnarray}
	\underline{\underline{M}} 
	& \equiv & 	
	\left[{ \begin{array}{cc}
		\underline{\underline M}^s & 
		\underline{\underline 0} \\
		\alpha 
		\underline{\underline M}^r 
		\underline{\underline B}^T & 
		\underline{\underline M}^r
	\end{array} }\right],
	\\
	\underline{\underline{K}} 
	& \equiv & 	
	\left[ \begin{array}{cc}
		\underline{\underline K}^s & 
		-\alpha 
		\underline{\underline B} \, 
		\underline{\underline K}^r \\
		\underline{\underline 0} & 
		\underline{\underline K}^r
	\end{array} \right],
	\\
	\underline{\underline{R}}
	 & \equiv & 	
	\left[ \begin{array}{cc}
		\underline{\underline I} & 
		-\alpha \underline{\underline B} \\
		\underline{\underline 0} & 
		\underline{\underline I}
	\end{array} \right],
\end{eqnarray}
and for convenience we have transformed to mass weighted coordinates 
$\underline{y}$ with the matrix $\underline{\underline{\gamma}}$.  We can 
rewrite Eq.~(\ref{eqn:eom_form}) as:
\begin{equation}
	\underline{\ddot{y}}(t) + 
	\underline{\underline{X}} \, 
	\underline{y}(t) 
	= 
	\underline{\underline{\gamma}}^{-1} 
	\underline{\underline{M}}^{-1} 
	\underline{\underline{R}} \, 
	\underline{F}(t),
	\label{eqn:eom_form2}
\end{equation}
where we have defined
\begin{equation}
	\underline{\underline{X}} \equiv 
	\underline{\underline{\gamma}}^{-1} 
	\underline{\underline{M}}^{-1}
	\underline{\underline{K}} \, 
	\underline{\underline{\gamma}}.
	\label{eqn:X_matrix}
\end{equation}
We may diagonalize $\underline{\underline{X}}$ using the transformation 
$\underline{\underline{D}} = \underline{\underline{U}}^{-1} 
\underline{\underline{X}} \, \underline{\underline{U}}$.  We now define our 
normal coordinates as $\underline{\eta}(t) \equiv 
\underline{\underline{U}}^{-1} \underline{y}(t)$.  For convenience, we also 
define a transformation matrix $\underline{\underline{V}} \equiv 
\underline{\underline{\gamma}} \, \underline{\underline{U}}$.  Substituting 
these into Eq.~(\ref{eqn:eom_form2}) and multiplying the entire expression 
by $\underline{\underline{U}}^{-1}$ we find
\begin{equation}
	\underline{\ddot{\eta}}(t) + 
	\underline{\underline{D}} \, 
	\underline{\eta}(t) 
	= 
	\underline{\underline{V}}^{-1} 
	\underline{\underline{M}}^{-1} 
	\underline{\underline{R}} \, 	
	\underline{F}(t).
	\label{eqn:nc_eom}
\end{equation}

The problem has now been reduced to $5+J$ decoupled harmonic oscillator 
equations.  To solve them we begin by taking the Fourier transform of 
Eq.~(\ref{eqn:nc_eom})
\begin{equation}
	\underline{\underline{G}}^{-1}(\omega) 
	\underline{\eta}(\omega)
	= 
	\underline{\underline{V}}^{-1} 
	\underline{\underline{M}}^{-1} 
	\underline{\underline{R}} \, 
	\underline{F}(\omega),
\end{equation}
where we have defined
\begin{equation}
	\underline{\underline{G}}^{-1}(\omega) 
	\equiv
	\underline{\underline D} -
	\omega^2 
	\underline{\underline I}.
\end{equation}
Because $\underline{\underline{D}}$ is diagonal, 
$\underline{\underline{G}}^{-1}(\omega)$ is also diagonal, so its inverse 
is just the diagonal elements inverted.  We can now easily solve for the 
normal coordinates:
\begin{equation}
	\underline{\eta}(\omega) 
	= 
	\underline{\underline{G}}(\omega) 
	\underline{\underline{V}}^{-1} 
	\underline{\underline{M}}^{-1} 
	\underline{\underline{R}} \, 
	\underline{F}(\omega).
	\label{eqn:nc_solution}
\end{equation}
To return to the original coordinates we reverse the transformations:
\begin{eqnarray}
   \left[\begin{array}{c}
      \underline{a}(\omega)  \\
      \underline{q}(\omega)
   \end{array} \right]
   & = &
   \underline{\underline{\gamma}} \, \underline{y}(\omega) \\
   & = &
   \underline{\underline{V}} \, \underline{\eta}(\omega)
   \label{eqn:V_eta} \\
   & = &
   \underline{\underline{V}} \, \underline{\underline{G}}(\omega) \,
	\underline{\underline{V}}^{-1} \underline{\underline{M}}^{-1} 
	\underline{\underline{R}} \, \underline{F}(\omega).
\end{eqnarray}
Note that Eq.~(\ref{eqn:V_eta}) provides a convenient way to transform to 
normal modes where the frequency response is simple.  The matrix 
$\underline{\underline{V}}$ is always invertible as we know the inverse of 
$\underline{\underline{U}}$ and $\underline{\underline{\gamma}}$ exist, 
thus making it possible to transform in both directions.  This 
transformation will be important in the final analysis of the detector  
discussed below.

\section{Mode channels}
\label{sec:mode_channels}

In our original TIGA model~\cite{Johnson_PRL_1993} we showed it was 
possible to combine the observable resonator displacements 
$\underline{q}(t)$ into a quantity which we called ``mode channels'' 
because they have a one-to-one correspondence with the quadrupole modes of 
a sphere, and thus the spherical amplitudes of a gravitational wave.  It is 
desirable at this point to develop a general expression for the equivalent 
of mode channels for any number and arrangement of radial resonant 
transducers.

We begin by taking the Fourier transform of Eqs.~(\ref{eqn:resonator_eom}) 
and (\ref{eqn:coupled_eom}):
\begin{equation}
	\left[
	\underline{\underline{K}}^r - 
	\omega^2 \underline{\underline{M}}^r
	\right] 
	\underline{q}(\omega) -
	\alpha \omega^2 
	\underline{\underline{M}}^r 
	\underline{\underline{B}}^T 
	\underline{a}(\omega)
	= 
	\underline{F}^r(\omega)
	\label{eqn:resonator_eom_ft2}
\end{equation}
\begin{equation}
	\left[
	\underline{\underline{K}}^s - 
	\omega^2
	\underline{\underline{M}}^s
	\right] 
	\underline{a}(\omega) -
	\alpha 
	\underline{\underline{B}} \, 
	\underline{\underline{K}}^r 
	\underline{q}(\omega)
	=
	-\alpha 
	\underline{\underline{B}} \, 
	\underline{F}^r(\omega) + 
	\underline{F}^s(\omega)
	\label{eqn:coupled_eom_ft2}
\end{equation}
For the moment, we are only interested in the force of the gravitational 
wave acting on the sphere, so we will assume to have a high signal-to-noise 
ratio, thus we can ignore the external forces on the resonators and set 
$\underline{F}^r(\omega) = 0$.  We can now solve for 
$\underline{F}^s(\omega)$ in terms of $\underline{q}(\omega)$
\begin{eqnarray}
	\underline{F}^s(\omega)
	& = &
	\left[
	\frac{1}{\alpha \omega^2} 
	\underline{\underline{K}}^s
	\left(
	\underline{\underline{B}} \, 
	\underline{\underline{M}}^r 
	\underline{\underline{B}}^T 
	\right)^{-1}
	\underline{\underline{B}} \, 
	\underline{\underline{K}}^r 
	\right.
	-
	\frac{1}{\alpha} 
	\underline{\underline{K}}^s
	\left(
	\underline{\underline{B}} \, 
	\underline{\underline{M}}^r 
	\underline{\underline{B}}^T 
	\right)^{-1}
	\underline{\underline{B}} \, 
	\underline{\underline{M}}^r \nonumber \\
	& & \left. - 
	\frac{1}{\alpha} 
	\underline{\underline{M}}^s
	\left(
	\underline{\underline{B}} \, 
	\underline{\underline{M}}^r 
	\underline{\underline{B}}^T 
	\right)^{-1}
	\underline{\underline{B}} \, 
	\underline{\underline{K}}^r
	+
	\frac{\omega^2}{\alpha} 
	\underline{\underline{M}}^s
	\left(
	\underline{\underline{B}} \, 
	\underline{\underline{M}}^r 
	\underline{\underline{B}}^T 
	\right)^{-1}
	\underline{\underline{B}} \, 
	\underline{\underline{M}}^r
	- 
	\alpha 
	\underline{\underline{B}} \, 
	\underline{\underline{K}}^r
	\right] 
	\underline{q}(\omega)
	\label{eqn:soln_F_full}
\end{eqnarray}

Eq.~(\ref{eqn:soln_F_full}) gives us the means, using the observable 
resonator displacements $\underline{q}$, to infer the force on the 
quadrupole modes applied by a gravitational wave.  However, the complicated 
frequency response will make its implementation difficult because all the 
parameters on the right hand side must be known.  While it may be possible 
to determine all of these parameters (see the Appendix for an example) we 
would prefer a technique that is not so strongly dependent upon measuring 
these.  In addition, we would prefer an arrangement where the frequency 
response can be simplified.

In the following sections we propose an alternative technique that does not 
require one to know all the parameters of the detector to high accuracy.  
Using a special symmetric arrangement of resonators, the above equations 
can be simplified.  Along with a special procedure, possibly unique to this 
arrangement, we can obtain all the information about the external forces 
without knowing all the parameters of the system to high accuracy.

\section{Direction finding technique}
\label{sec:direction_finding}

\subsection{General technique}

It is very desirable to demonstrate a general algorithm for finding the 
location of an arbitrary excitation solely from the mode channel 
amplitudes.  We were able to find and successfully test such an algorithm, 
one suggested by the ellipsoidal picture for the shape of the modes 
discussed above.

The measured amplitudes of the quadrupole modes directly tell us the 
relative amounts of each of the 5 basis ellipsoids that must be 
superimposed to get the net ellipsoidal deformation.  We denote the 5 
ellipsoidal amplitudes by $h_m$, and call them the vibration amplitudes in the 
``spherical representation.''

We can also define a matrix of the quadratic form $h_{ij}$ whose elements 
form a complete set of amplitudes in what we call the ``cartesian 
representation.'' The connection between representations is easily found to 
be
\begin{eqnarray}
   h_{ij}(t) 
   & = & 
   \left[\begin{array}{ccc}
      h_{xx} & h_{xy} &  h_{xz} \\
      h_{yx} & h_{yy} &  h_{yz} \\
      h_{zx} & h_{zy} &  h_{zz}
   \end{array} \right] \nonumber \\
   & = &
   \left[\begin{array}{ccc}
      h_{1} - \frac{1}{\sqrt{3}}h_{5} & h_{2} &  h_{4} \\
      h_{2} &  -h_{1} - \frac{1}{\sqrt{3}}h_{5} &  h_{3} \\
      h_{4} &  h_{3} &  \frac{2}{\sqrt{3}} h_{5}
   \end{array} \right].
   \label{eqn:cartesian_strain_tensor}
\end{eqnarray}  
The connection between this representation and the geometry of ellipsoids 
comes again from the principal axes theorem, which states that the three 
eigenvectors of the matrix $h_{ij}$ are parallel to the three principal 
axes, and the radial deviations $dr_i$ are the corresponding eigenvalues of 
$h_{ij}$.

An excitation can be classified by the shape and orientation of the 
ellipsoid it produces.  Once this shape is realized, one need only solve 
for the eigenvalues and eigenvectors of the matrix $h_{ij}$ to determine 
the direction of the excitation.  The exact interpretation of the 
eigenvalues and eigenvectors will of course depend upon the expected 
ellipsoidal deformation.

\subsection{Inferring a gravitational wave's direction}

The method described in the previous section can be used to determine the 
direction of an incident gravitational wave.  As shown in 
Sec.~\ref{sec:quadrupole_gravity}, the gravitational field can also be 
represented by an ellipsoid derived from the electric components of the 
Riemann tensor~\cite{Eardley_PRL_1973,Eardley_PRD_1973}.  It describes the 
relative acceleration of gravity that causes an ellipsoidal deformation of 
an initially spherical group of free test particles.  In a conventional 
gauge, that ellipsoid is also described by the 9 spatial components of the 
gravitational strain tensor $h_{ij}$ in cartesian coordinates.  It can 
easily be shown that this tensor is in exact one-to-one correspondence to 
the cartesian amplitudes of vibration of the sphere, so in this paper we 
have used the same symbol $h_{ij}$ for both.

Now, the direction problem in gravitation requires only knowledge of what 
sort of ellipsoid is produced by a gravitational wave.  By examining the 
conventional description of the strain tensor of a wave according to 
General Relativity~\cite{MTW}, we find that one principal axis of the 
ellipsoid is aligned with the direction of propagation, and that the 
corresponding radial deviation is zero.  Therefore, we need only determine 
the wave's ellipsoid, and then we know the eigenvector of the zero 
eigenvalue points at the source.  (This position determination is unique 
only within a hemisphere; sources in diametrically opposite directions are 
indistinguishable.)

Note that this method does not require intensive calculations, such as 
those used to compute the maximum likelihood estimates performed by Zhou and 
Michelson~\cite{Zhou_PRD_1995}; however, its effectiveness in the presence 
of noise still needs to be evaluated.

\subsection{Inferring the direction of a radial impulse}

Since a laboratory source of gravitational waves does not exist, we need an 
alternative type of excitation for testing this technique.  We find that a 
radial impulse to the surface of a sphere is a good substitute.

We present here a simple picture of the antenna's response to a radial 
impulse.  Suppose a sphere has a degenerate quadrupole mode multiplet, so 
we are free to choose a basis set with {\it arbitrary\/} orientation to 
describe it.  If we choose an orientation with the $z'\text{-axis}$ of the 
mode frame to be along the direction of the impulse, then only a {\it 
single\/} mode ($Y_5$) in that frame will be excited (all of the other 
modes have a vanishing radial component of their eigenfunctions at this 
location, which makes their ``overlap'' integral with the impulse vanish).  
The corresponding ellipsoid produced is an oblate spheroid which has 
maximum radial deviation at the location of the impulse, and two half-size 
radial deviations of opposite sign in the orthogonal directions.  
Therefore, the location of the impulse is given by the eigenvector with the 
largest eigenvalue.  This also provides a check for the assumed shape (in a 
measurement with a high signal-to-noise ratio) as the other two eigenvalues 
should be equal to each other, but half the size and opposite in sign of 
the first.

\section{The ideal truncated icosahedral arrangement}
\label{sec:ideal_TI}

\subsection{Symmetry}

When we began this problem \cite{Johnson_PRL_1993} we introduced a special 
arrangement of 6 resonators which we termed a Truncated Icosahedral 
Gravitational Wave Antenna (TIGA) shown in Fig.~\ref{fig:TI_arrangement}.  
We proposed using a truncated icosahedron as an approximation to a sphere, 
however the only requirement for the Truncated Icosahedral (TI) arrangement 
was that the resonant transducers be placed at positions on the surface of 
a sphere at the center of six non-antipodal pentagon faces of an imaginary 
truncated icosahedron (or dodecahedron) concentric to the sphere.

The original TIGA model \cite{Merkowitz_PRD_1995} assumed perfect symmetry 
of the sphere as well as the tuning and placement of the resonant 
transducers.  While the effects of deviations from perfect symmetry on a 
sphere's uncoupled quadrupole modes have been studied 
\cite{Lobo_EPL_1996,Merkowitz_PRD_1996,Coccia_PLA_1996}, we still need to 
investigate the effect of asymmetries on our ability to properly interpret 
the signals from resonant motion sensors.  This is why we have kept the 
equations general until now.  It is possible to investigate alternative 
arrangements of radial resonators, such as the one proposed by Lobo and 
Serrano \cite{Lobo_EPL_1996}, with the above framework, however, we will 
limit ourselves here to the TI arrangement for reasons stated in the 
introduction.

The symmetry of a TI, shown in Fig.~\ref{fig:TI_symmetry}, greatly 
simplifies various aspects of the problem; not only in the calculations 
that follow, but also in the construction of such a device.  A TI has 32 
flat surfaces suitable for mounting transducers, calibrators, balancing 
weights and suspension attachments.  In addition, the symmetry makes 
machining the solid TI relatively simple \cite{Merkowitz_Thesis}.  However, 
as stated above, the only requirement is on the placement of the 
resonators, not on the shape of the ``spherical'' mass.

The high symmetry of the TI arrangement becomes apparent when you examine 
its pattern matrix $\underline{\underline{B}}$.  Each pattern vector is 
orthogonal to the others, and each has the same magnitude, 
$\sqrt{\frac{3}{2\pi}}$, or in other words:
\begin{equation}
  \underline{\underline B} \, 
  \underline{\underline B}^T 
  = 
  \frac{3}{2\pi} 
  \underline{\underline I}.
  \label{eqn:BB=I}
\end{equation}
This property causes the cross terms between sphere modes in the normal 
mode eigenfunctions to vanish.  In addition to the orthogonality, the sum 
of the components of each pattern vector vanishes:
\begin{equation}
   \underline{\underline B}\,
   \underline{1}
   =
   \underline{0}.
   \label{eqn:B1=0}
\end{equation}
The $6 \times 1$ column vector $\underline{1}$ is defined to have all 
elements equal to unity, while the $5 \times 1$ column vector 
$\underline{0}$ has all elements equal to zero.

\subsection{Eigenfunction solution}

The symmetry of the pattern matrix also suggested that there might be an 
analytic solution for the collection of eigenvectors 
$\underline{\underline{U}}$ and the eigenvalue matrix 
$\underline{\underline{D}}$ of Eq.~(\ref{eqn:nc_eom}).  Examination of the 
numerical results suggested a likely form for $\underline{\underline{U}}$, 
and substitution in the equations verified that it was a solution and 
determined the values of the constants.  The details of this solution can 
be found elsewhere \cite{Merkowitz_PRD_1995,Merkowitz_Thesis}.

It is convenient to divide the resulting set of eigenvectors, 
$\underline{\underline{U}}$, into three groups.  The first two groups each 
contain 5 column eigenvectors and we denote them by 
$\underline{\underline{U}}_+$ and $\underline{\underline{U}}_-$:
\begin{equation}
   \underline{\underline{U}}_\pm = 
   n_\pm 
   \left[ \begin{array}{c}
      \underline{\underline{I}}  \\
      c_\pm \underline{\underline{B}}^T
   \end{array} \right].
   \label{eqn:U_pm}
\end{equation}
The physical interpretation of these is simple: each coupled eigenmode 
``mimics'' the motion of one of the uncoupled sphere eigenmodes.  In other 
words, each coupled resonator's radial motion is proportional to the 
uncoupled sphere eigenfunctions at that resonator's location.  This 
amplified version of a mode's pattern vector is either in-phase and 
down-shifted in frequency, or anti-phase and up-shifted in frequency.  The 
frequency shifts are all identical, so that the quintuplet of degenerate 
bare sphere-modes has bifurcated into up-shifted and down-shifted 
degenerate quintuplets of modes.  The amount of frequency shifting is given 
by the eigenvalues of $\underline{\underline{U}}$, which are the diagonal 
elements of the matrix $\underline{\underline{D}}$.  The identity matrix in 
Eq.~(\ref{eqn:U_pm}) is an indication that energy will {\it not\/} be 
transferred from one sphere mode to another.  The $\pm$ notation has been 
used on the dimensionless constants $n_\pm$ and $c_\pm$ as well to refer to 
the up ($+$) or down ($-$) shifting of the frequencies.

The remaining single eigenvector is
\begin{equation}
   \underline U_o = 
   \left[ \begin{array}{c}
      \underline{0} \\
      n_o \underline{1}
   \end{array} \right].
   \label{eqn:U_o}
\end{equation}
This mode is at the original sphere frequency and does not strongly 
interact with a gravitational wave.  All the resonators move in unison and 
the sphere modes do not move at all.

\subsection{Ideal mode channels}

Now let us see what the mode channels look like for the ideal TI 
arrangement:
\begin{eqnarray*}
	\underline{\underline{M}}^s = m_s \underline{\underline{I}} &, \;&
	\underline{\underline{M}}^r = m_r \underline{\underline{I}} ,  \\
	\underline{\underline{K}}^s = k_s \underline{\underline{I}} &, \;&
	\underline{\underline{K}}^r = k_r \underline{\underline{I}},
\end{eqnarray*}
thus
\begin{equation}
	\underline{F}^s(\omega)
	=
	\left[
	\frac{2 \pi}{3 \alpha \omega^2 m_r} 
	\left(k_s - \omega^2 m_s \right)
	\left(k_r - \omega^2 m_r \right) 
	- 
	\alpha k_r \right] 
	\underline{\underline{B}} \, 
	\underline{q}(\omega).
	\label{eqn:ideal_mc}
\end{equation}
What is striking about Eq.~(\ref{eqn:ideal_mc}) is that all the complicated 
frequency dependence has been separated from the matrices and a simple 
linear combination of the resonator responses can be made to obtain all the 
information about the external forces.  We, therefore, define a quantity 
$\underline{g}$ which does not contain the complicated frequency 
dependence, but still maintains the one-to-one correspondence with the 
quadrupole components of the external force acting on the sphere:
\begin{equation}
	\underline{g} \equiv \underline{\underline{B}}\,\underline{q}.
	\label{eqn:original_mc}
\end{equation}
The components of the mode channels $g_m$ can be used as a substitute for 
the amplitudes $h_m$ in order to solve for the directional information of 
an excitation, such as a gravitational wave.  A practical application of 
this technique for an impulse excitation to the surface of a prototype 
antenna will be described later in Sec.~\ref{sec:uncoupled_prototype}.

\subsection{Resonator ellipsoids}

From the equations of motion of an ideal TIGA, we found that the 
eigenfunctions of the coupled modes were such that the motion of the 
resonators mimicked the ellipsoidal deformation of the sphere's surface 
either in phase or anti-phase.  We, therefore, can picture the collective 
motion of the six resonators to describe six ``resonator ellipsoids'', 5 of 
which are mimicking the ``quadrupole ellipsoids'' of the sphere.  The sixth 
resonator ellipsoid is just a sphere, where the six resonators are moving 
in unison with equal amplitude and phase, and the sphere surface does not 
move at all, as described by Eq.~(\ref{eqn:U_o}).

Each individual resonator now represents a superposition of the point radial 
deformation of the 6 resonator ellipsoids at a particular position.  The 
transformation between point radial deformations $\underline{q}$ and 
ellipsoidal amplitudes $\underline{g}$ is given by the pattern matrix 
$\underline{\underline{B}}$ defined by the positions of the resonators and 
the orientation of the 5 quadrupole ellipsoids relative to a fixed lab 
frame:
\begin{equation}
	\underline{g} = \underline{\underline{B}} \; \underline{q}.
	\label{eqn:pattern_matrix}
\end{equation}
Note that Eq.~(\ref{eqn:pattern_matrix}) is identical to 
Eq.~(\ref{eqn:original_mc}) for transforming to mode channels.  Through 
this discovery we realize that in the case of the TI arrangement, we can 
think of mode channels as the result of a linear coordinate transformation 
from resonator displacements to ellipsoidal deformations.  This 
relationship is {\it not\/} general!  In other arrangements of resonators, 
the equivalent resonator ellipsoids do not, in general, mimic the 
quadrupole ellipsoids, thus are not the same as mode channels.  To produce 
mode channels for other arrangements, one would have to introduce the 
complicated frequency response described by Eq.~(\ref{eqn:soln_F_full}).

\section{A nearly truncated icosahedral arrangement}
\label{sec:real_TI}

We have seen how simple things become when the TI arrangement is used, but 
what happens if the system is not ideal?  Using a numerical model, 
described below, we investigated the effects on the eigenfunctions due to 
perturbations of the system parameters.  We found that small deviations of 
the various parameters (of the order 1\%) from the ideal TI case did not 
significantly change the resonator ellipsoids.  We therefore will discuss a 
situation where the tolerance on the parameters is relaxed to be of the 
order a few percent.  In an actual experiment this is a rather poor level of 
precision; one expects to be able to do much better.

\subsection{Normal mode coordinates}

While all the signal information is contained in the resonator 
ellipsoids, it is useful to be able to transform the data to normal 
mode coordinates using Eq.~(\ref{eqn:V_eta}) where the frequency 
response is simple.  While not important for transforming between 
point radial coordinates to ellipsoidal coordinates, the symmetry 
breaking can be significant when transforming to normal mode 
coordinates.

To overcome this, we developed an {\it in situ\/} measurement 
technique \cite{Merkowitz_OMNI_1997} to determine the transformation 
matrix $\underline{\underline{V}}$.  The transformation matrix can be 
measured by applying a continuous sinusoidal force anywhere on the 
sphere's surface at the frequency of one of the normal modes.  (Note 
that this technique requires the normal modes to be {\it 
non-degenerate\/}, thus it is actually preferable to have a small 
amount of symmetry breaking.) The frequency response of the resonators 
will be simple because they are being driven at a single frequency.  
From Eq.~(\ref{eqn:V_eta}) we see that the amplitude and phase of 
their response make up a single column of $\underline{\underline{V}}$.  
By exciting each normal mode in turn, the complete 
$\underline{\underline{V}}$ matrix can be measured.  The only 
assumption made in calculating $\underline{\underline{V}}$ is that the 
resonators are ``close'' to the TI arrangement so that an ideal 
pattern matrix $\underline{\underline{B}}$ can be used and the 
quadrupole ellipsoids can be replaced by 
Eq.~(\ref{eqn:pattern_matrix}).

Once $\underline{\underline{V}}$ is measured the antenna can be 
operated to observe gravitational waves.  The response of the 
resonators can be recorded and transformed to normal mode coordinates:
\begin{equation}
 	\underline{\eta} = \underline{\underline{V}}^{-1} \left[\left[
 	\begin{array}{c}
 		\underline{\underline{B}}Ê\; \underline{q}_{-}  \\
 		\underline{q}_{-}
 	\end{array}
 	\right] + \left[
	\begin{array}{c}
 		-\underline{\underline{B}} \; \underline{q}_{+}  \\
 		\underline{q}_{+}
 	\end{array}
	\right] 	\right].
	\label{eqn:nm_transformation}
\end{equation}
Note that the resonator response must be bandpass filtered to separate 
the in-phase ($-$) and anti-phase ($+$) resonator ellipsoids.  Since 
the frequency response of the normal modes is simple they can easily 
be fit for various parameters such as phase and amplitude.  This 
information can then be transformed to mode channels using using 
Eq.~(\ref{eqn:V_eta}).  From the mode channels the direction and 
amplitude information of a possible gravitational wave event can be 
calculated as described above.

\subsection{Numerical simulation of errors}

\subsubsection{Parameters}

Now that we have written down the solutions to the equations of motion we 
can look at what are the parameters of the system and how uncertainties and 
deviations of them will affect a measurement.  For the TIGA case we have 
the following parameters:
\begin{displaymath}
	\begin{array}{ccll}
	6 & \times & k^r_j              & \text{resonator spring constants} \\
	5 & \times & k^s_m              & \text{sphere mode spring constants} \\
	6 & \times & \phi_j             & \text{resonator positions}  \\
	6 & \times & \theta_j           & \text{resonator positions}  \\
	5 & \times & \beta_m            & \text{sphere mode orientations}  \\
	5 & \times & \gamma_m           & \text{sphere mode orientations} \\
	6 & \times & m^r_j              & \text{resonator masses} \\
	5 & \times & m^s_m              & \text{quadrupole mode masses} \\
	6 & \times & \epsilon^r_j       & \text{resonator radial couplings} \\
	6 & \times & \epsilon^\theta_j  & \text{resonator transverse couplings} \\
	6 & \times & \epsilon^\varphi_j & \text{resonator transverse couplings} \\
	\hline
	62 & \multicolumn{3}{l}{\text{total parameters.}}
	\end{array}
\end{displaymath}

The angles $\beta_m$ and $\gamma_m$ describe the orientation of the 
quadrupole modes relative to a fixed lab frame, and will be discussed 
further below.

The parameters $\epsilon^r_j$, $\epsilon^\theta_j$, and 
$\epsilon^\varphi_j$ are a measure of the coupling of the transducers to 
the radial and transverse motion of the sphere surface.  In the above model 
we assumed $\epsilon^r_j=1$ and $\epsilon^\theta_j=\epsilon^\varphi_j=0$ 
because we felt their inclusion was unnecessary as transducers are 
currently available that do not strongly couple to transverse motion.  
However, it is useful to test this assumption and determine how strong a 
requirement should be set for the actual instrument.  We include them here 
by replacing the pattern matrix defined by 
Eq.~(\ref{eqn:pattern_matrix_def}) with
\begin{equation}
	B_{mj} \equiv
	\frac{1}{\alpha} \left(
	  \epsilon^r_j \hat{\bbox{r}}
	+ \epsilon^\theta_j \hat{\bbox{\theta}}
	+ \epsilon^\varphi_j \hat{\bbox{\varphi}} 
	\right) \cdot \bbox{\Psi}_m\left({\theta_j,\phi_j}\right)
\end{equation}
This should give a good approximation of the effects of transverse 
coupling, without the need of changing the model of the resonators from one 
dimensional harmonic oscillators.  We considered these parameters as 
independent from each other.  One might relate them with a parameter such 
as the angle between the transducer axis and the normal to the sphere 
surface, however we do not do this because for an actual resonant 
transducer there are several other mechanisms that can lead to transverse 
coupling, thus keeping these parameters independent seems reasonable.

Some of the above parameters can potentially be measured directly, such as 
the masses, however, we include them here for generality.  In addition, the 
5 quadrupole mode masses $m^s_m$ would normally be set equal to each other 
and to the physical mass of the sphere, however, again for generality we 
kept it as a parameter.  In the appendix we describe a procedure to measure 
most of these parameters, however the resonator ellipsoid method described 
above makes this unnecessary.

\subsubsection{Simulation results}

The transformation matrix $\underline{\underline{V}}$ can be measured to 
very high accuracies, but our assumption that the resonator ellipsoids 
still mimic the sphere ellipsoids will have some error associated with it.  
This error will propagate through the analysis and into the results of a 
measurement.  We, therefore, studied the effects of small perturbations to 
the above parameters on our ability to accurately determine the direction 
of an excitation.

We developed a Monte Carlo type simulation where we added a small random 
perturbation (uniform distribution) to the above parameters within a 
specified tolerance.  We then simulated an excitation and calculated the 
direction using the resonator ellipsoid method.  The direction calculation 
assumed the ideal case: it was not given knowledge of the true values of 
the parameters.

As shown in Fig.~\ref{fig:direction_error}, the results of the numerical 
simulation indicate that a direction calculation becomes unreliable only 
after the tolerance of all the parameters exceeds about 3\%.  This is 
certainly an obtainable level of precision.  Fig.~\ref{fig:tiga_error_sang} 
shows the solid angle estimation error 
$\Delta\Omega$~\cite{Gursel_PRD_1989} for several tolerances.  We also 
varied the location of the excitation, but found no significant difference 
in the results.  Note that these are systematic errors due to the analyses 
technique, not random errors as the figures might imply.

To put these results into perspective, we compared these systematic errors 
to the random error due to a finite signal-to-noise ratio as calculated by 
Zhou and Michelson~\cite{Zhou_PRD_1995}.  We find that one would need a 
signal-to-noise ratio of about 1000 in energy before our systematic errors 
become significant (choosing a tolerance of 2\%).  While one might hope to 
observe sources at this level, the most optimistic predictions lead to 
considerably smaller signal-to-noise ratios~\cite{Finn_OMNI_1997}.  We, 
therefore, feel that the systematic errors are sufficiently low that there 
is no need to develop an alternative technique that requires precise 
knowledge of the parameters.

Looking at the individual contribution to the errors from each parameter we 
determined which were most dominant.  We found that the most significant 
errors came from the resonator positions $\phi_j$ and $\theta_j$.  All the 
other parameters had associated errors at least one order of magnitude in 
$\Delta\Omega$ lower than those from the resonator positions for reasonable 
tolerance levels.  This included the errors associated with the coupling 
parameters $\epsilon^r_j$, $\epsilon^\theta_j$, and $\epsilon^\varphi_j$, 
thus justifying our earlier decision to omit them from the model.

Perturbations to the sphere mode orientations $\beta_m$ and $\gamma_m$ did 
not lead to {\it any\/} errors!  We expected that simple linear combination 
of the quadrupole modes do not lead to any error, however, what is 
surprising is that a direction calculation's ignorance of the true 
orientation in a non-degenerate system does not lead to any errors.  This 
is important as it tells us that it is unnecessary to measure the mode 
orientation before equipping the sphere with resonant transducers (as was 
done on the prototype TIGA for other reasons discussed below).  While these 
parameters may not completely describe the effects of deviations of the 
quadrupole modes from an ideal sphere, we found from 
measurements~\cite{Merkowitz_PRD_1996} that they are the dominant effect of 
symmetry breaking.  The fact that they do not contribute {\it at all\/} to 
the errors on a measurement also frees us from putting strong constraints 
on the spherical mass.  This allows us, for example, to put a hole through 
the center of the sphere for suspension purposes, or use a TI (or some 
other ``spherical'' shape) instead of a sphere.

\section{The LSU prototype TIGA}
\label{sec:prototype}

The above model outlines a clear algorithm for obtaining the gravitational 
amplitudes from a spherical antenna.  However, like most models, we must 
evaluate its worth with an actual experiment.  We therefore constructed a 
room temperature prototype TIGA.  In the following sections we describe how 
the prototype was used to: first, verify that a TI has the same mode 
structure as a sphere; second, determine the effects of asymmetries on the 
sphere modes, such as a hole drilled through the center for suspension; 
third, verify the mode channel and ellipsoidal theories; and finally, 
verify the direction finding algorithms.

The prototype TI was machined from a bar of aluminum alloy 6063 that had 
previously been used as a cylindrical gravitational wave detector and was 
known to have good mechanical properties.  Some key dimensions of the TI 
are shown in Fig.~\ref{fig:TI_shop_drawing}.  The prototype had a center of 
mass suspension.  A hole was bored along a diameter that started from the 
center of a hexagon face.  The hole changed diameter just above the center 
of mass, and a thin titanium suspension rod, which widened to a cone at one 
end to mate with the hole's change in diameter, was inserted from the large 
diameter side.

The prototype was first suspended and tested without mechanical resonators 
attached.  This testing gave many insights into the differences between an 
ideal sphere and a real one.  The results of this testing was summarized 
elsewhere \cite{Merkowitz_PRD_1996} but we include  here some of the 
important results that are needed to describe the coupled system.

Once the uncoupled tests were completed, resonant transducers were 
attached, and the coupled system was studied.  Preliminary results of these 
tests were also summarized elsewhere 
\cite{Merkowitz_OMNI_1997,Merkowitz_PRL_1997}, but we report here the 
completed work in detail.

While we have attempted to report here as many of the important aspects of 
the experiment as possible, we have omitted many of the specifics of this 
particular apparatus; for those details we refer the reader to 
reference~\cite{Merkowitz_Thesis}.

\section{The uncoupled prototype}
\label{sec:uncoupled_prototype}

\subsection{Normal Mode Frequencies }

The measured frequency spectrum of the uncoupled TI is shown in 
reference~\cite{Merkowitz_PRD_1996}.  We were able to identify most of the 
predominant modes using solutions to the elastic equations of a 
sphere~\cite{Lobo_PRD_1995} and a finite element model of a 
TI~\cite{Merkowitz_Thesis}.  We found the degeneracy of all multiplets to 
be lifted by a small amount: 1\% or less in frequency.  We were able to 
match the measured frequencies of most of the multiplets to the theory for 
a sphere to better than 1\%.

The modes of most interest for gravitational wave detection are the 5 
members of the lowest quadrupole mode multiplet near 3235 Hz.  For a 
homogeneous isotropic sphere, those modes are exactly degenerate.  We found 
that this quintuplet was split into two doublets and a singlet, spread over 
a range of 0.8\% in frequency, as shown in Fig.~\ref{fig:quadrupole_spec}.  
Additional data (not shown), confirmed that the two peaks labeled as 
doublets are each composed of two modes split by about 1 Hz.

Upon reflection, we realized that the suspension hole bored through the TI 
must be the primary cause for the splitting of the quintuplet.  It breaks 
the spherical symmetry, but preserves cylindrical symmetry about the hole 
axis.  The specific identification of the multiplets shown in 
Fig.~\ref{fig:quadrupole_spec} was surmised on physical grounds, and 
confirmed by measurements described below.  We have not attempted to 
calculate the magnitude of the splitting caused by the hole, so we cannot 
make a comparison with the data, however, this effect has subsequently been 
confirmed by others~\cite{Coccia_PLA_1996}.

\subsection{Monopole Mode Calibration}
\label{sec:calibration}

This experiment dealt with high signal-to-noise ratios, and absolute energy 
calibration was unnecessary.  However, it was important to know the 
relative sensitivity of the motion sensors and correct for any differences.  
The monopole, or breathing, mode of a sphere (which for this TI had a 
frequency near 6880~Hz) is a spherically symmetric radial expansion and 
contraction of the surface.  The TI had no other modes close in frequency 
to the monopole mode.  This made it ideal to measure the relative 
sensitivity of the motion sensors.

We excited vibrations of the TI with radial impulses from a hammer at 
various locations on the surface.  Shown in Fig.~\ref{fig:monopole}, we 
found that the responses of the six motion sensors, at the monopole 
frequency, were identical in phase, and independent of the position of the 
impulse, but differed systematically in amplitude by up to 10\%.  These 
amplitude differences were due to the quality of the attachment of the 
motion sensors as well as gain differences in the electronics chain.  These 
measured gain deviations were then used to correct the amplitudes in all 
the subsequent measurements.  This method proved to be very convenient as 
the motion sensors did not have to be removed or remounted, which was found 
to change their sensitivity.

\subsection{Simple mode channels}

We observed the quadrupole mode multiplet by sampling the motion of the TI 
at 6 discrete positions, using small, non-resonant, accelerometers waxed to 
the surface in the TI arrangement as shown in 
Fig.~\ref{fig:TI_arrangement}.  According to the standard normal mode 
picture of vibrational mechanics, the free motion at these points, or any 
point on the surface, can be viewed as the combination, or superposition, 
of the response of the normal modes.  Thus each motion sensor will record a 
different linear superposition of the responses of all the modes.

A hammer was used to impulsively excite vibrations of the sphere.  
Narrow-band filtering was used to isolate the quadrupole modes.  The 
measured response of each motion sensor is shown in 
reference~\cite{Merkowitz_PRD_1996}.  As expected, the non-degeneracy of 
the modes caused the individual sensor outputs to display the various modes 
beating against each other, making it difficult to make a {\it direct\/} 
estimate of the amplitude of each normal mode.

We showed above that the desired mode amplitudes could be separated out by 
combining the outputs of all the sensors in special linear combinations, 
whose coefficients were grouped together into the pattern matrix 
$\underline{\underline{B}}$.  We called these combinations ``mode 
channels'' to indicate they had a one-to-one correspondence with the 
quadrupole normal mode amplitudes of the sphere.  For the case of the 
uncoupled prototype, we do not need to included the measurement of the 
matrix $\underline{\underline{V}}$ to convert to normal modes, as the 
uncoupled sphere quadrupole modes {\it are\/} the normal modes, thus their 
frequency response is simple.  In addition, for the case of {\it this 
prototype\/}, we could not use this procedure because several of the modes 
were nearly degenerate, thus exciting them individually with a simple 
sine-wave excitation was impossible.

To obtain nearly perfect mode channels, we rotated the spherical harmonics 
that determined the pattern matrix, until we found the best fit to a single 
frequency in each mode channel.  We chose to use the y-convention for the 
Euler angles \cite{Goldstein_1980} to perform the rotations.  The rotation 
$\alpha$ about the $z\text{-axis}$ was not used because it had little 
effect on the fit.  The $\beta$ rotation about the new $y\text{-axis}$ 
mixed mode 5 with the other 4 modes, while maintaining orthogonality.  The 
$\gamma$ rotation about the new $z\text{-axis}$ mixed the new modes 1 with 
2, and 3 with 4, but not 1 with 3 or 4, etc.  Therefore, these rotation 
angles could be different for the two pairs and still maintain 
orthogonality.  Mode 5 was unaffected by any $\gamma$ rotation.  The best 
fit values for the rotation angles from the lab coordinate system shown in 
Fig.~\ref{fig:TI_arrangement} were $\gamma_{12} = -0.1$, $\gamma_{34} = 
-7.2$, and $\beta = 1.0$.

Each mode channel was well separated from the others and behaved as 
expected, an exponentially decaying sine wave.  By examining the power 
spectrum, we determined that the residual amplitude of the ``wrong'' modes 
present in a channel was less than 2\%.  This small residual admixture may, 
or may not, be due to imprecise positioning of the accelerometers.

\subsection{Simple impulse test}

As a final test of the uncoupled system, we applied several radial impulses 
to the center of nine different faces of the TI, and then calculated the 
locations from the algorithm described above.  The results of this 
comparison are shown in reference~\cite{Merkowitz_PRD_1996}.  The locations 
calculated from the data were very consistent; with three hits at each 
location, the overall standard deviation from the mean was $\sim 
0.4^\circ$.  The calculated locations were all within $\sim 3\%$ of the 
values expected from the measured geometric position of the impulse hammer.  
The deviation from the expected values is apparently a systematic error, 
perhaps from imprecise placement of the accelerometers or the impulse 
hammer.  Below we describe the results of a similar test, but with the 
accelerometers replaced by resonant transducers.

\subsection{Insights from the uncoupled prototype}

Experiments on the uncoupled prototype showed that the departures from 
perfect spherical behavior and symmetry were not large.  The quadrupole 
modes were no longer degenerate, but the source of their frequency 
splitting is understood.  The eigenfunctions of the uncoupled TI were found 
to be unchanged in shape from those of a perfect sphere by an amount less 
than 2\%; the main effect of the symmetry breaking was to fix them in a 
particular orientation.  The simple impulse test confirmed the practicality 
of the direction finding technique.  From these results, we conclude that a 
TI represents a good approximation for a sphere and is sufficient for use 
as an omnidirectional gravitational wave antenna.  Knowing these results, 
we were confident enough to instrument the prototype TI with resonant 
transducers to fully test the TIGA theory.

\section{The prototype with resonant transducers}
\label{sec:coupled_prototype}

\subsection{Transducer design and attachment}

Section~\ref{sec:sphere_with_res} lists the rudimentary requirements for a 
resonator, but practical considerations require a more extensive list.  
First, the ``transducer mode'' must be reasonably easy to tune to the 
quadrupole frequency.  Second, the transducer mode must be purely radial, 
so that it couples strongly only to the radial motion of the quadrupole 
modes.  Third, there should not be any other modes of the resonator nearby 
in frequency.  Fourth, there must be a practical method of attachment with 
sufficient mechanical~Q.

The design we adopted for the prototype, shown in Fig.~\ref{fig:fathead}, 
approximates a lumped mass and a spring.  The lumped mass, or ``head'', is 
attached to a thin stem, or ``neck.'' The neck is fixed to a base which is 
then attached to the surface of the prototype.  These three parts were 
machined from a single piece of aluminum.  The transducer mode is such that 
radial motion of the head compresses and extends the neck against the base.  
While the neck is relatively rigid in the radial direction, it is 
relatively flexible in the transverse directions, which decouples the 
transducer mode from transverse motions.  While designing the resonator, 
the length and diameter of the neck can be adjusted to move the rocking and 
torsional modes of the resonator well below the transducer mode frequency.

A piezoelectric strain gauge was epoxied to the neck of each resonator.  
The strain induced in the crystal will be proportional to the change in 
length of the neck, thus providing an efficient way of observing the motion 
of the resonators.  The output of the strain gauges were first demodulated 
to low frequency using 6 separate lock-in amplifiers, using the same 
reference, and then recorded on a high speed data acquisition system.  The 
resonators and measurement system are described in detail in reference 
\cite{Merkowitz_Thesis}.

We used finite element analysis to fix the final parameters of the 
resonator.  The final tuning of the resonators was done while attached to 
the TI.  The equations of motion for this system are given in 
Sec.~\ref{sec:sphere_with_res} taking the number of resonators equal to 
one.  We measured the coupled mode frequencies of the prototype and one 
resonator and compared them to the eigenvalue solution of 
Eq.~(\ref{eqn:eom_matrix}) to determine the spring constant of the small 
resonator.  We had two practical options for tuning the resonator: reduce 
the mass of the head, or lower the spring constant by reducing the diameter 
of the neck.

We attached the resonators to the prototype with epoxy.  While this method 
may not lead to the best mechanical Q, we found it was sufficient.  The 
coupled modes had a Q of about $10^3$ in vacuum while the uncoupled sphere 
modes had a Q of about $10^4$.  We suspect this difference is due to the 
method of attachment.  In air the coupled modes had a very poor Q, thus we 
felt it was necessary to perform all test of the coupled system under 
vacuum.

The resonators were attached to the prototype one at a time, and the 
frequencies of the coupled modes were measured after each change.  The 
calculated and measured quadrupole mode frequencies were fairly consistent 
with each other.  The non-degeneracy of the prototype's quadrupole modes 
did not introduce much deviation from a perfectly degenerate system.  It 
was also found that neither the toroidal modes nor the monopole mode of the 
sphere were shifted by more than 1 Hz when the resonators were added.  
Fig.~\ref{fig:coupled_data} shows the results of the frequency measurements 
of the coupled modes for each addition of a resonator.  The results are 
compared with what is expected from the eigenvalue solution of 
Eq.~(\ref{eqn:eom_matrix}) beginning with the measured uncoupled 
eigenfrequencies.  The two sets are consistent within 0.2\%.  While we 
consider this very good agreement, Lobo and Serrano found slightly better 
agreement (possibly due to better numerical precision), with this data, by 
applying the equations of motion in an equivalent, but different 
form~\cite{Lobo_EPL_1996}.

\subsection{Transformation to normal modes}

With the 6 resonators attached, we were ready to observe the sphere modes 
using the resonator ellipsoid technique.  The first step was to measure the 
transfer function $\underline{\underline{V}}$.  We attached a simple 
non-resonant piezoelectric shaker to the surface of the prototype.  The 
frequencies of the 11 normal modes were accurately measured by driving the 
shaker with a single frequency sine-wave, adjusting the frequency until a 
maximum response of a single normal mode was observed.  With this system in 
equilibrium, we recorded the response of the 6 resonators to the continuous 
wave excitation.  The frequency of the excitation was then changed to 
measure the next normal mode frequency and the above steps repeated.

As shown in Fig.~\ref{fig:cw_raw}, the frequency response of the 
resonators was simple because we were driving at a single frequency.  The 
amplitudes of their response made up a single column of 
$\underline{\underline{V}}$.  By exciting each normal mode in turn, the 
complete transformation matrix was measured.  We repeated this measurement 
for several different locations of the shaker, all of which gave consistent 
results.

Shown in Fig.~\ref{fig:cw_nm} is the results of using the measured 
transformation matrix $\underline{\underline{V}}$ to calculate the response 
of the normal modes to the normal mode excitation of Fig.~\ref{fig:cw_raw}.  
For this case the TI was driven with a continuous wave excitation at the 
frequency of the fifth normal mode.  As shown in the figure, only the 
fifth normal mode was excited, as expected.  Again, this experiment had 
high signal-to-noise ratios, thus the essentially flat lines of the 
non-excited modes actually represent a ``leakage'' level of about 5\% in 
amplitude.

\subsection{Impulse direction test}

To combine the entire TIGA technique into a single test, we applied an 
impulse excitation to the surface of the prototype TIGA to determine if the 
location of the impulse can be measured from the response of the 
resonators.  An impulsive force was applied to the surface of the TI by 
sending a short electrical pulse to a non-resonant piezoelectric shaker 
attached to the surface.

Shown in Fig.~\ref{fig:burst_raw} is a typical response of the six 
transducers to an impulsive excitation.  Following the technique described 
above, this data can be transformed to normal coordinates using the 
matrix $\underline{\underline{V}}$.  The results of this transformation to 
the data of Fig.~\ref{fig:burst_raw} is shown in Fig.~\ref{fig:burst_nm}.
As expected, the data separated into 11 channels, each containing a single 
frequency representing the response of a single normal mode.  Since each 
channel contains only a single frequency, it is relatively easy to fit them 
for their phase and amplitude at the time of excitation.   Once these 
quantities are found we can transform them to mode channels and compute 
the location of the impulse as described above.

The results of this analysis for the various impulse locations is shown in 
Fig.~\ref{fig:burst_location}.  The locations calculated from the data were 
very consistent; with several impulses at each location, the overall 
standard deviation from the mean was $0.1^\circ$.  On average, the 
calculated locations were all within $2.7^\circ$ of the values expected 
from the geometrically measured position of the center of the shaker.

The deviation from the expected values can be accounted for by the accuracy 
of the excitation method.  The shaker used to apply the impulsive force did 
not actually apply a ``point'' impulse, but rather one that was distributed 
over a ring about the circumference of the shaker.  By systematically 
repositioning the shaker, we determined that the ``true'' location of the 
impulse was anywhere within $2.5^\circ$ of the geometric center of the 
shaker.  A more precise way of exciting the prototype would have been 
preferred, however, we found this method to be adequate to verify the 
principle of the technique.

\section{Summary}

Experiments on the prototype TIGA showed that the departures from ideal 
behavior were not large.  In every case, ways could be found to handle the 
asymmetries without major difficulty, and to some extent they actually 
simplify the problem.  A technique for determining the location of an 
external excitation, including that of a gravitational wave, from the 
motion sensor data was developed which, except for some bandpass filtering, 
is simply linear algebra.  This makes its implementation simple in an 
automated data analysis system.  The {\it in situ\/} measurement technique 
takes into account most deviations from perfect symmetry and the resulting 
transformation matrices enable the data to be transformed to a space where 
the frequency complications can be easily handled.  The algorithm was 
tested on the prototype TIGA and was found to be consistent with the 
measured results within the accuracy of the experiment.  Since all the 
techniques described can be applied {\it in situ\/}, they are directly 
applicable for use on a real spherical antenna searching for gravitational 
waves.  

\acknowledgments

We thank W.~O.~Hamilton for many years of essential advice and support on 
this project.  The final data analysis and writing of this paper was done 
at Eindhoven University of Technology and at the INFN Laboratori Nazionali 
di Frascati; S.~M.~M.~thanks A.~T.~A.~M.~de~Waele, E.~Coccia, G.~Pizzella 
and the rest of the GRAIL and ROG collaborations for their support during 
this time.  This research was supported by the National Science Foundation 
under Grant No.~\mbox{PHY-9311731}.

\appendix

\section*{Resonator channels}

To measure all the parameters of a spherical antenna {\it in situ\/} 
without introducing any new parameters we could do the following.  Excite 
one of the resonators with an external force and measure the response of 
all the resonators.  If we were to excite at some other location, this 
would introduce new unknown parameters such as the location of the excitor.

We assume that we have a high signal-to-noise ratio so that we can ignore 
any external forces acting directly on the sphere modes.  We therefore can 
set the forces $\underline{F}^s = 0$.  We now write 
Eqs.~(\ref{eqn:resonator_eom_ft2}) and~(\ref{eqn:coupled_eom_ft2}) as
\begin{equation}
	\left(
	\underline{\underline{K}}^r 
	- \omega^2 
	\underline{\underline{M}}^r
	\right)
	\underline{q}(\omega) 
	- \alpha \omega^2 
	\underline{\underline{M}}^r 
	\underline{\underline{B}}^T 
	\underline{a}(\omega)
	= 
	\underline{F}^r(\omega)
	\label{eqn:rc_resonator_eom}
\end{equation}
\begin{equation}
	\left(
	\underline{\underline{K}}^s 
	- \omega^2 
	\underline{\underline{M}}^s
	\right)
	\underline{a}(\omega)
	- \alpha 
	\underline{\underline{B}} \, 
	\underline{\underline{K}}^r 
	\underline{q}(\omega)
	=
	-\alpha \underline{\underline{B}} \, 
	\underline{F}^r(\omega).
	\label{eqn:rc_coupled_eom}
\end{equation}
Solving for $\underline{F}^r(\omega)$ in terms of the observable 
$\underline{q}(\omega)$ we find
\begin{equation}
	\underline{F}^r(\omega)
	=
	\left[
	\underline{\underline{I}}
	-\alpha^2 \omega^2 
	\underline{\underline{M}}^r 
	\underline{\underline{B}}^T
	\left(\underline{\underline{H}}^s(\omega)\right)^{-1}
	\underline{\underline{B}}
	\right]^{-1}
	\left[
	\underline{\underline{H}}^r(\omega)
	-\alpha^2 \omega^2 
	\underline{\underline{M}}^r 
	\underline{\underline{B}}^T
	\left(\underline{\underline{H}}^s(\omega)\right)^{-1}
	\underline{\underline{B}} \, 
	\underline{\underline{K}}^r
	\right]
	\underline{q}(\omega). 
	\label{eqn:resonator_channels}
\end{equation}
For the case of an ideal TIGA Eq.~(\ref{eqn:resonator_channels}) 
can be further simplified:
\begin{equation}
	\underline{F}^r(\omega)
	=
	\frac{
	\left(k_s - \omega^2 m_s \right) \left(k_r - \omega^2 m_r \right) -
	\frac{3}{2 \pi} \alpha^2 \omega^2 m_{r} k_{r}
	}{
	\left(k_s - \omega^2 m_s \right) - \frac{3}{2 \pi} \alpha^2 \omega^2 m_{r}
	}
	\underline{q}(\omega) 
\end{equation}

One can imagine performing this experiment and then fitting the resulting 
data for the various parameters.  However, during initial attempts to 
implement this technique we found the level of parameter fitting was 
complicated, even for advanced techniques such as simulated annealing, 
perhaps because global minimums did not exist.  While it may be possible to 
accurately fit for these parameters, we preferred to avoid such a task by 
developing and implementing the method of resonator ellipsoids discussed 
above.

\bibliographystyle{prsty}

\begin{figure}

\caption{The truncated icosahedral gravitational wave antenna (TIGA) with 
transducer locations indicated.  The transducers lie at two polar angles, 
$\theta=37.3773^\circ$ and $79.1876^\circ$.  Their azimuthal angles are 
multiples of $60^\circ$.}

\protect\label{fig:TI_arrangement}
\end{figure}

\begin{figure}

\caption{The symmetry of a Truncated Icosahedron}

\label{fig:TI_symmetry}
\end{figure}

\begin{figure}

\caption{The results of a numerical simulation of the systematic error on a 
source direction measurement due to a finite tolerance on the system 
parameters.  The simulated wave is linearly polarized with direction 
$\theta = 1$ rad, $\phi=2$ rad.  Each point represents a single direction 
measurement with the system parameters varied within the specified 
tolerance.}

\label{fig:direction_error}
\end{figure}

\begin{figure}

\caption{The solid angle direction estimation error as a function of the 
tolerance on the system parameters.  Each x represents the results of a 200 
trial simulation for a single source direction with all the system 
parameters varied within the corresponding tolerance.  Changing the source 
direction produced similar results.}

\label{fig:tiga_error_sang}
\end{figure}

\begin{figure}
\caption{Schematic of the prototype Truncated Icosahedron.}

\label{fig:TI_shop_drawing}
\end{figure}

\begin{figure}
\caption{The fine structure of the power spectrum of an impulse excitation 
of the TI's first quadrupole mode multiplet.  The five degenerate modes of 
an isotropic homogeneous sphere were split into two close doublets and a 
singlet.  Each mode is identified with its corresponding spherical 
harmonic.}

\label{fig:quadrupole_spec}
\end{figure}

\begin{figure}
\caption{The response of the six accelerometers to an impulse excitation of 
the monopole mode of the prototype TI.  Since the coupling to this mode 
is the same for all the accelerometers, it can be used to adjust for 
any differences in gain between the accelerometers and their readout 
electronics.}

\label{fig:monopole}
\end{figure}

\begin{figure}
\caption{Schematic of the resonant transducer.}

\label{fig:fathead}
\end{figure}

\begin{figure}
\caption{Frequency measurements of the coupled modes for each addition 
of a resonator.  The solid lines are the measured values and the 
dotted lines are the calculated.  The lines that are double in height 
represent degenerate doublets.}

\label{fig:coupled_data}
\end{figure}

\begin{figure}
\caption{A typical response of the six resonators to a continuous wave 
excitation applied to the surface of the prototype TI.  The excitation for 
this case was at the frequency of the fifth normal mode.  The outputs of 
the six resonant transducers was demodulated using 6 separate lock-in 
amplifiers using the same reference frequency at 3235 Hz (for clarity, only 
the in-phase is plotted).}

\label{fig:cw_raw}
\end{figure}

\begin{figure}
\caption{The response of the 11 normal modes to a continuous wave 
excitation at the frequency of the fifth normal mode calculated from the 
data shown in Fig.~\protect\ref{fig:cw_raw}.  As expected, only the fifth 
normal mode shows a large response.}

\label{fig:cw_nm}
\end{figure}

\begin{figure}
\caption{A typical response of the six resonant transducers to an impulsive 
excitation applied to the surface of the prototype TI at time $t=0$.  The 
outputs of the six resonant transducers was demodulated using 6 separate 
lock-in amplifiers using the same reference frequency at 3235 Hz (for 
clarity, only the in-phase is plotted).  The irregular response indicates 
that several normal modes contribute to the motion.}

\label{fig:burst_raw}
\end{figure}

\begin{figure}
\caption{The response of the 11 normal modes to an impulse excitation 
applied to the surface of the prototype TI at time $t=0$, calculated by a 
linear combination of the data shown in Fig.~\protect\ref{fig:burst_raw}.  
The regular response indicates that each channel corresponds to a single 
normal mode.}

\label{fig:burst_nm}
\end{figure}

\begin{figure}
\caption{Location of several impulses applied to the prototype TIGA.  The 
x's mark the locations calculated from the motion sensor data, and the 
nearby o's mark the location of the center of the shaker measured 
geometrically.}

\label{fig:burst_location}
\end{figure}

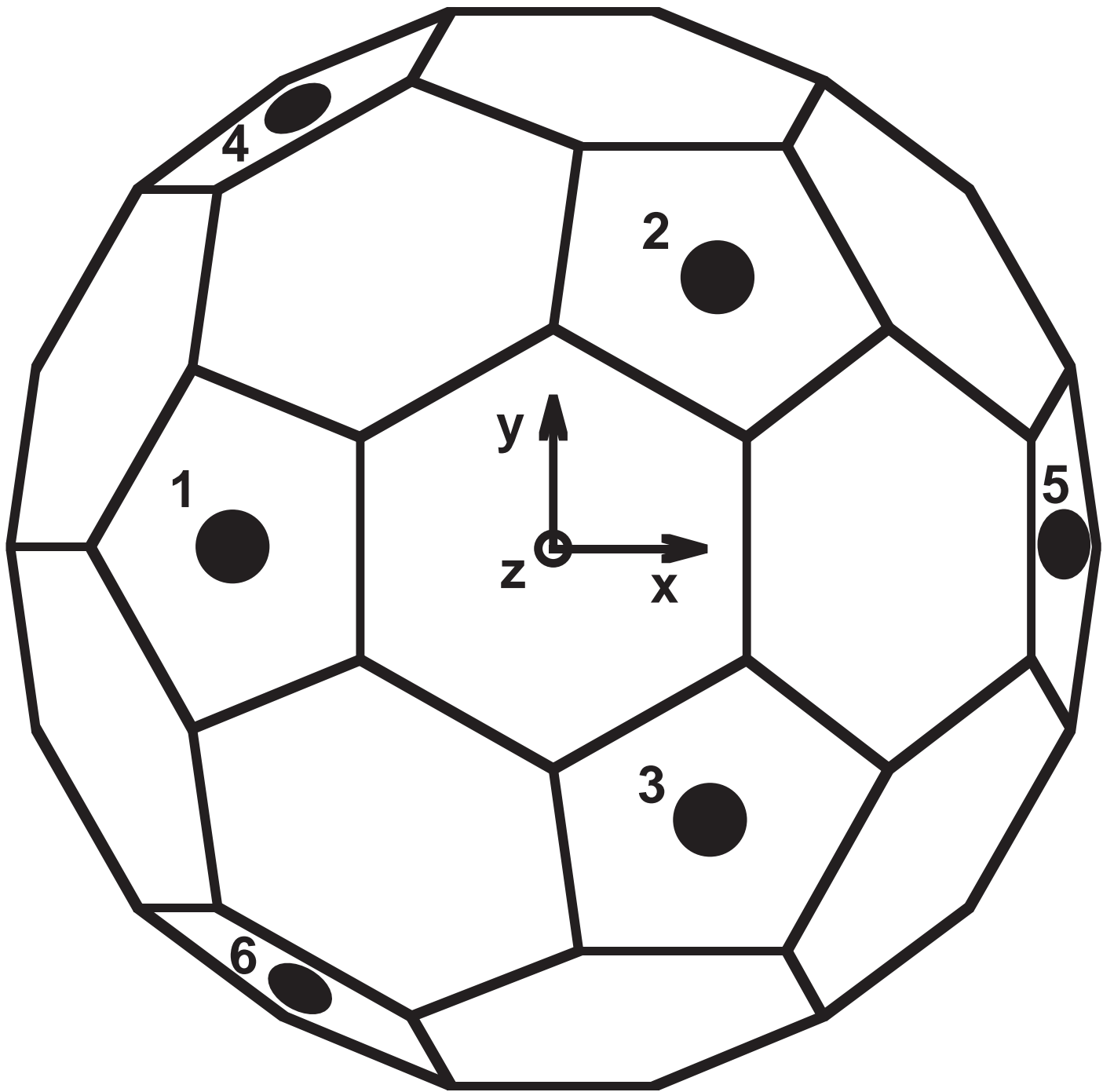
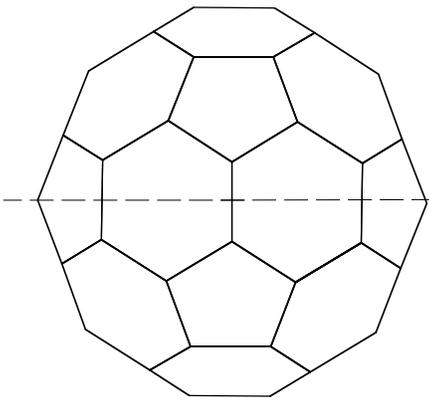
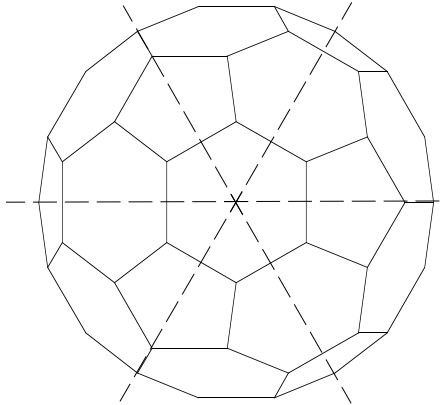
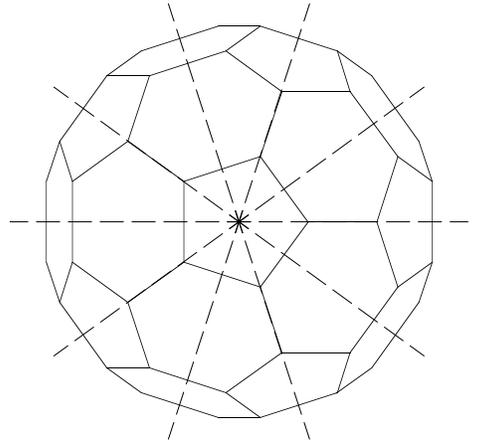
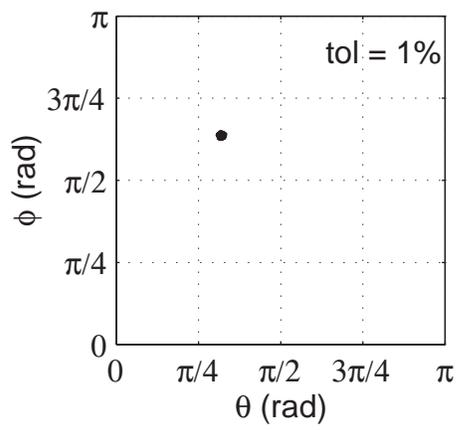
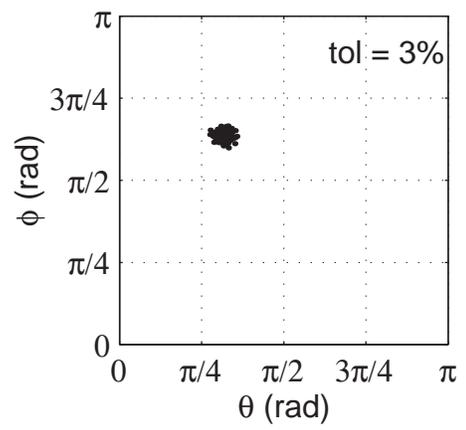
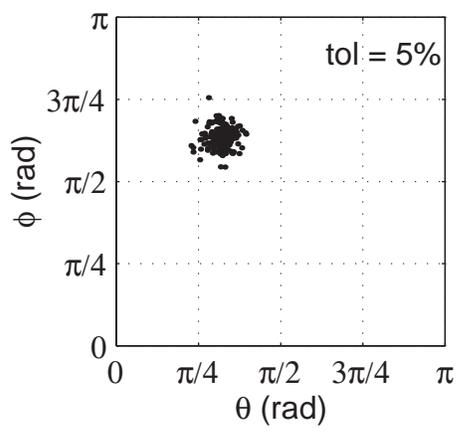
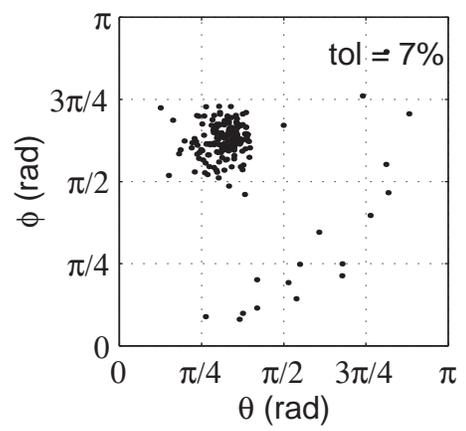
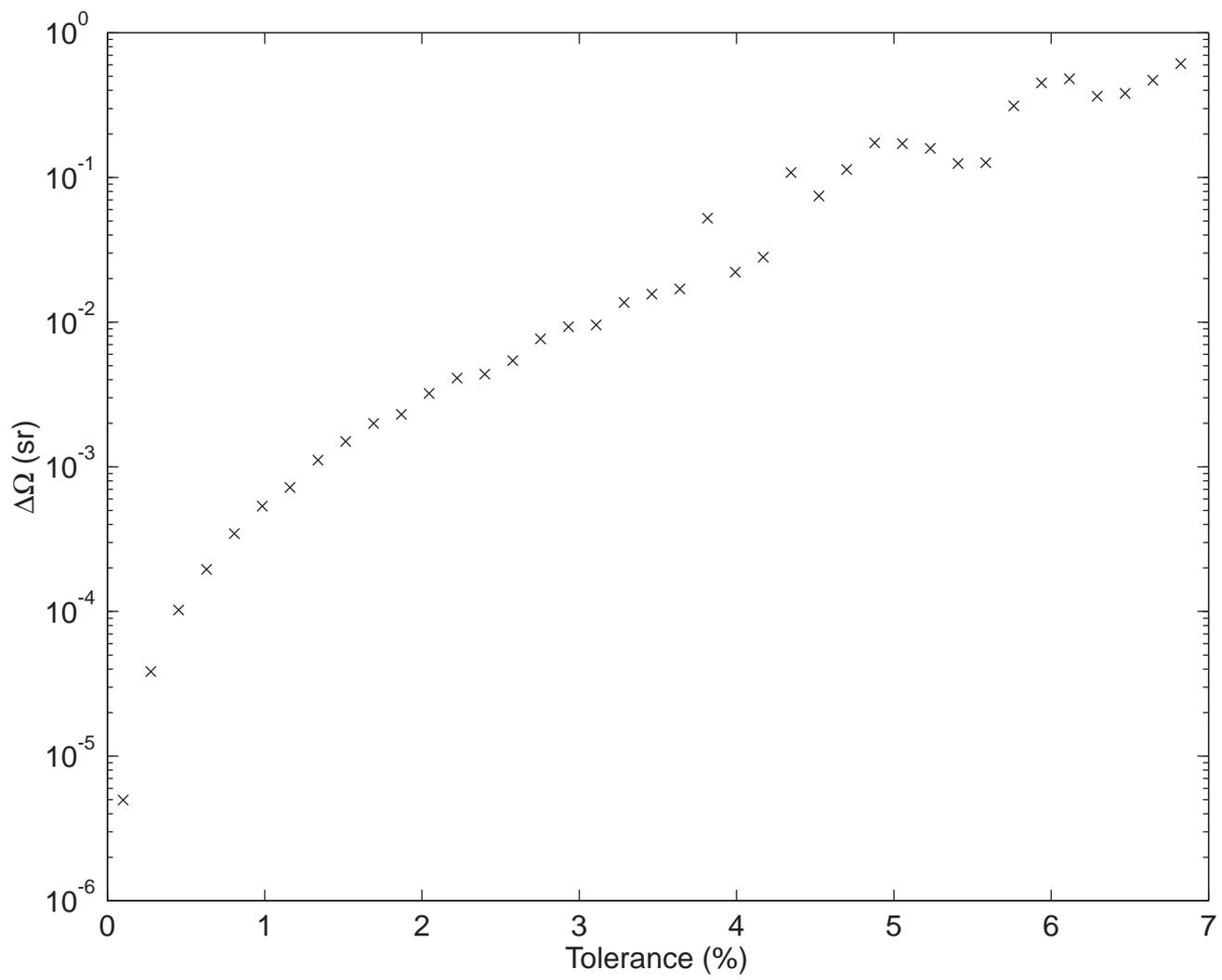
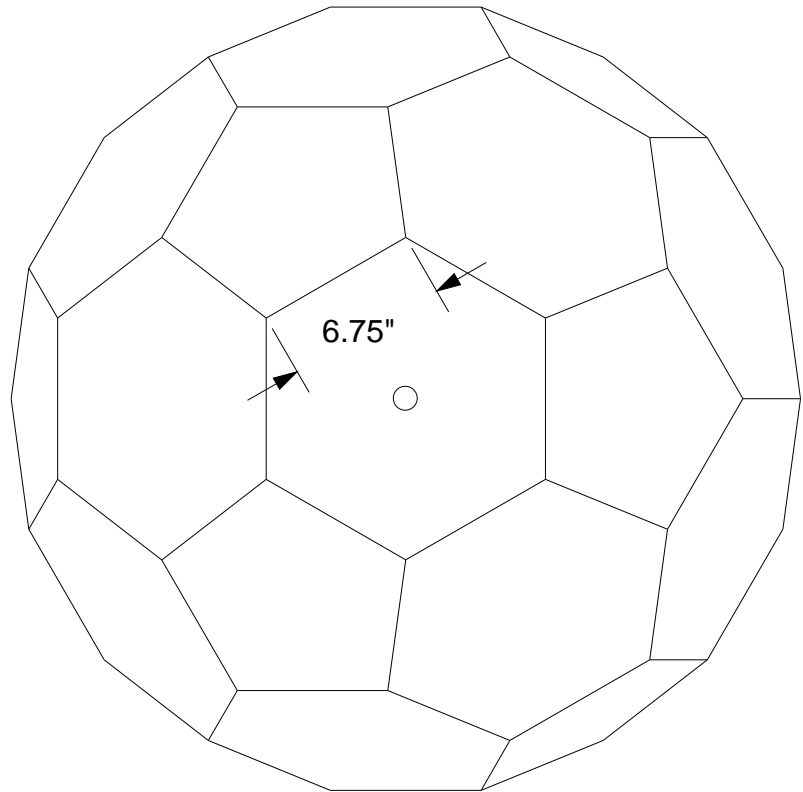
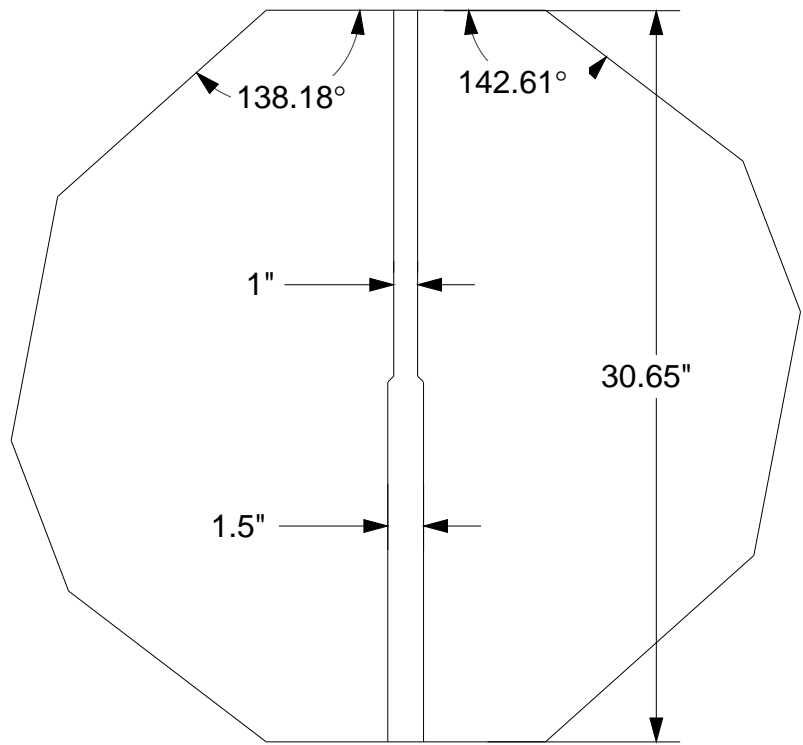
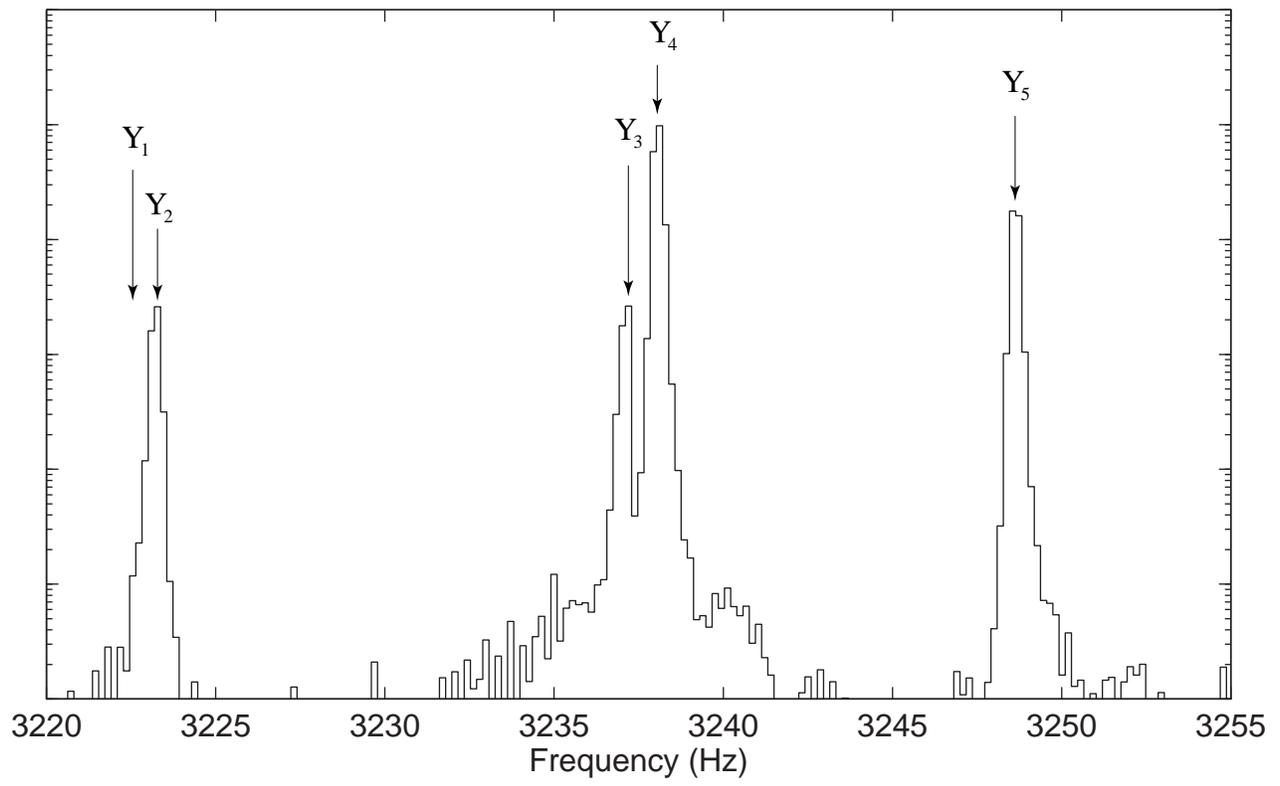
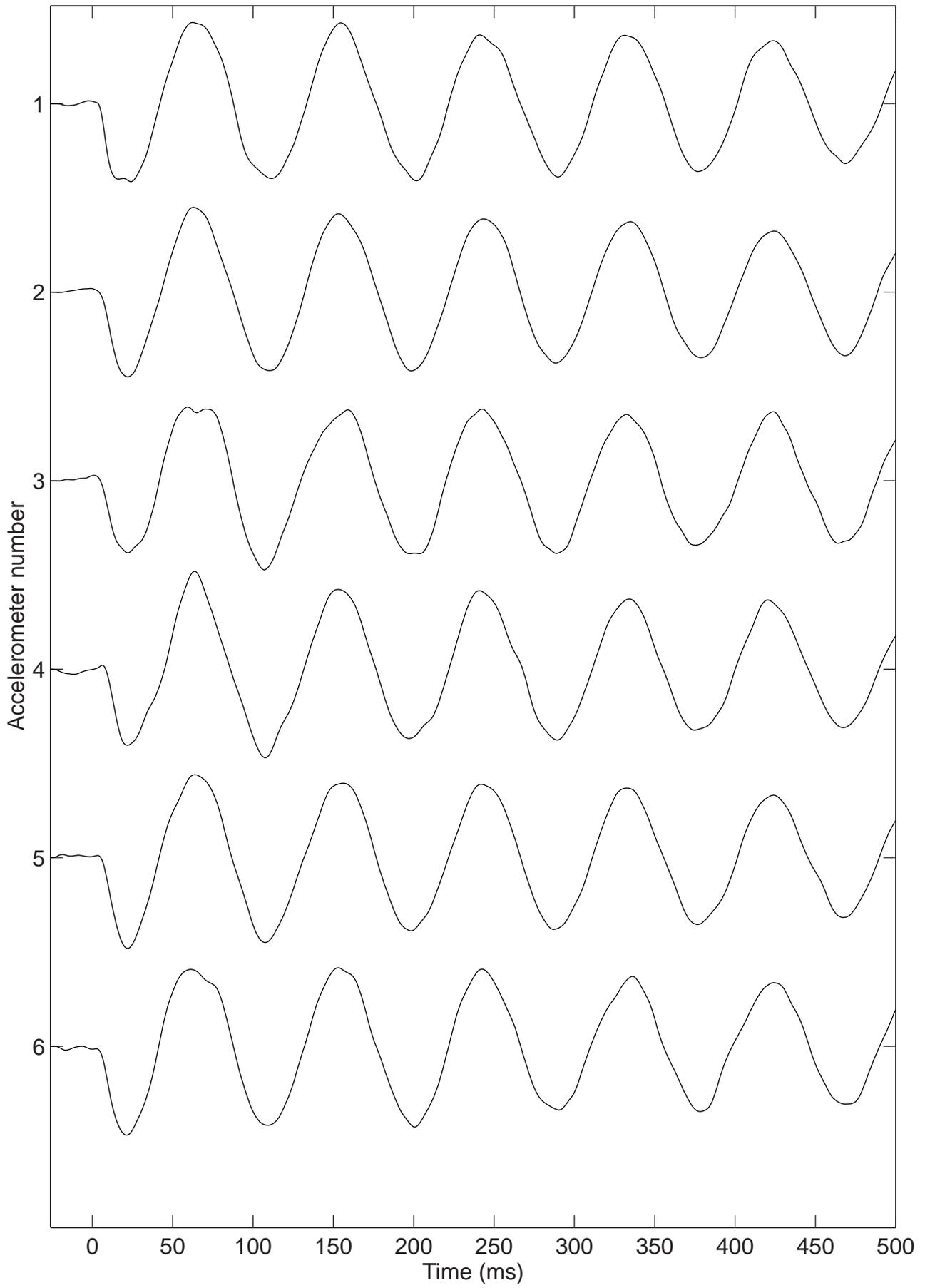
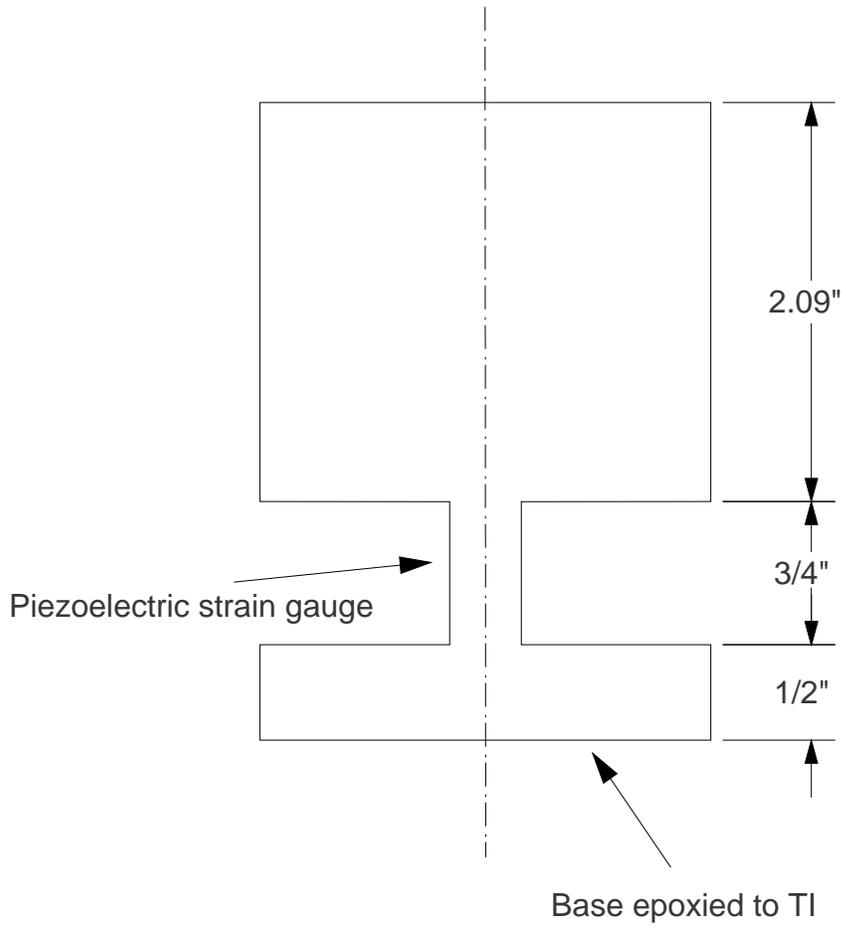
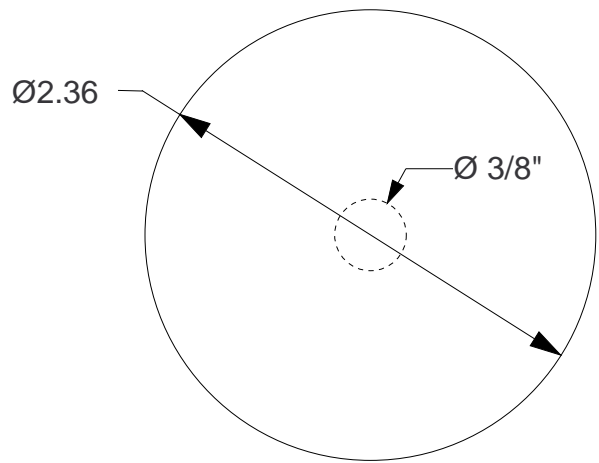
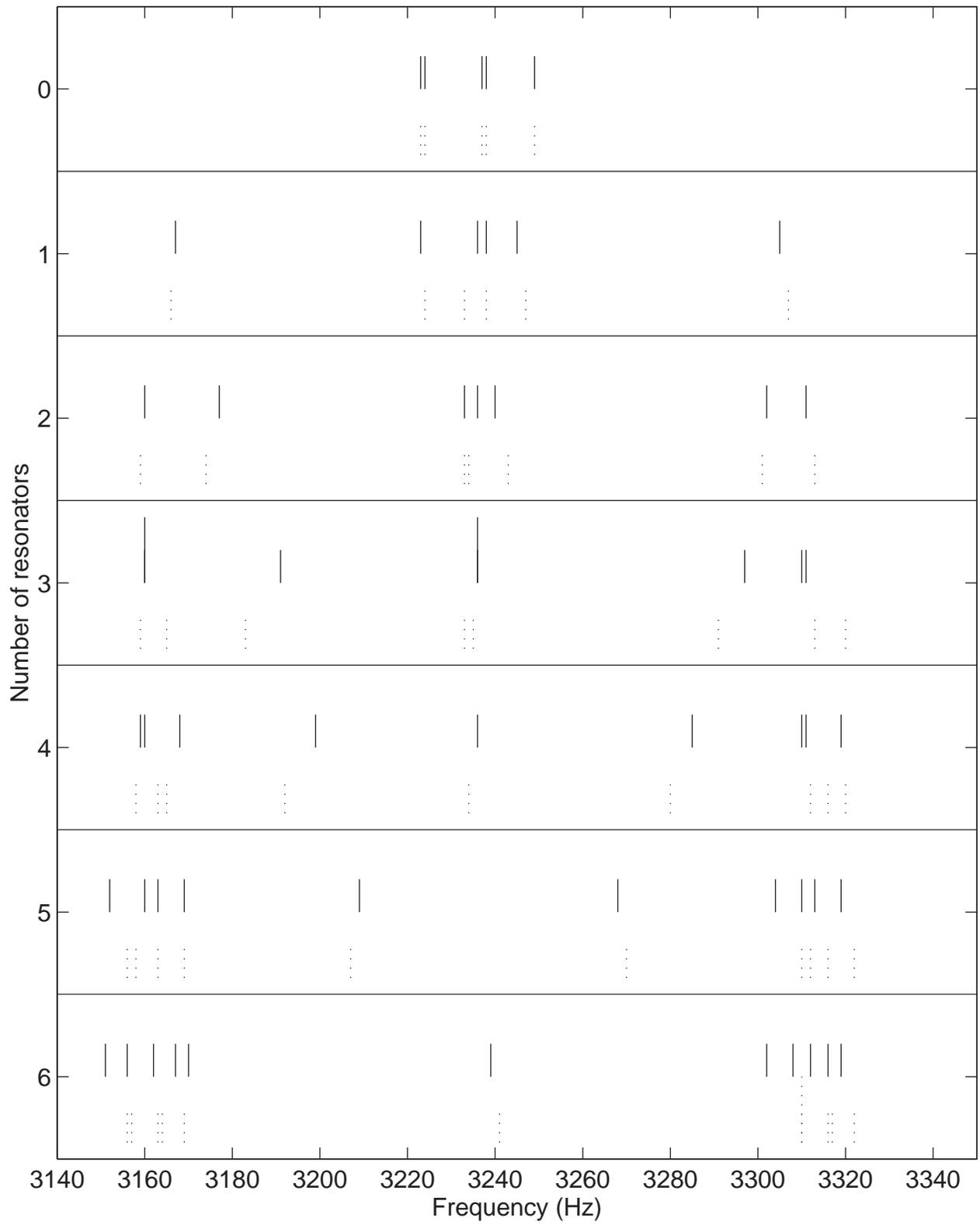
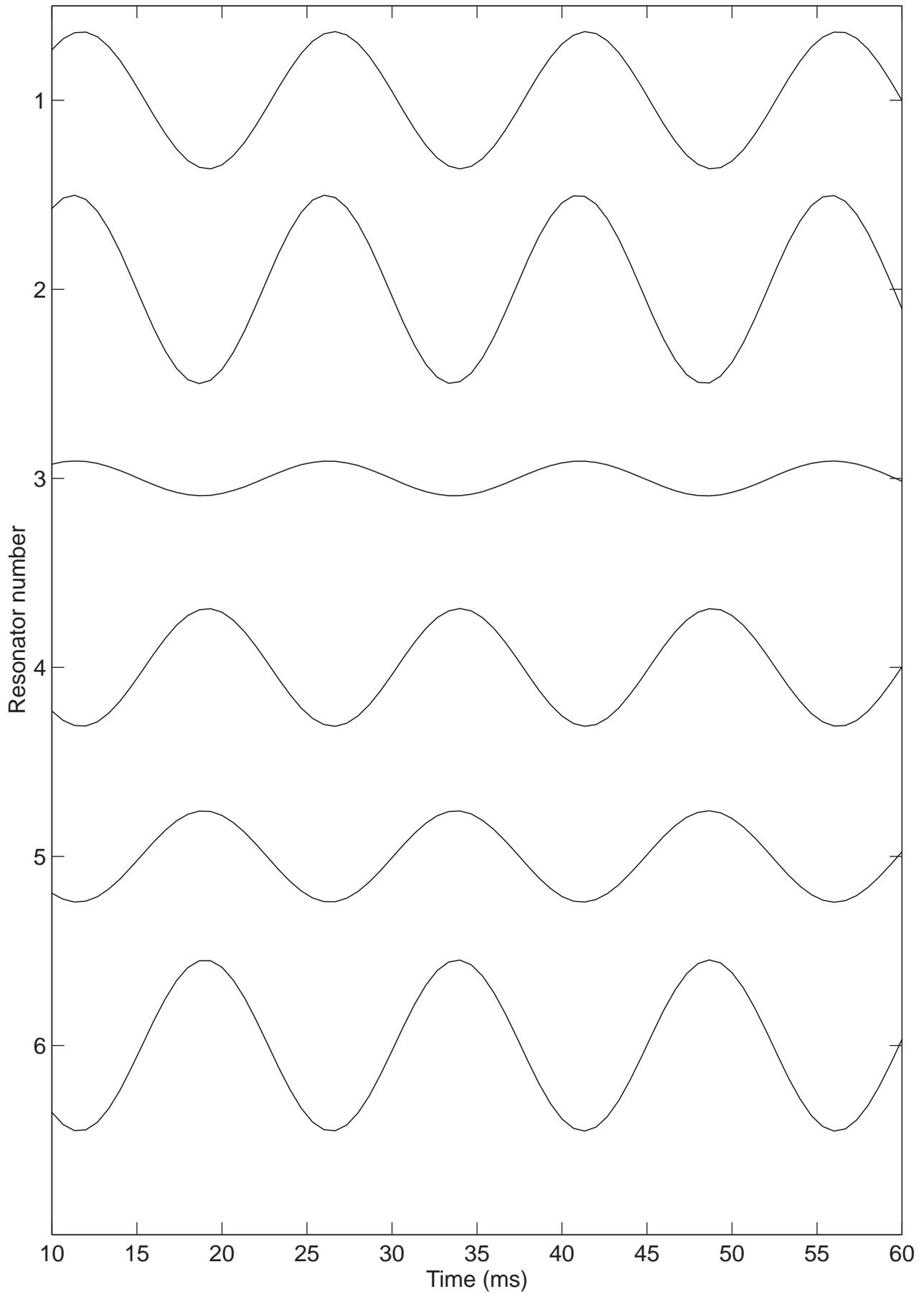
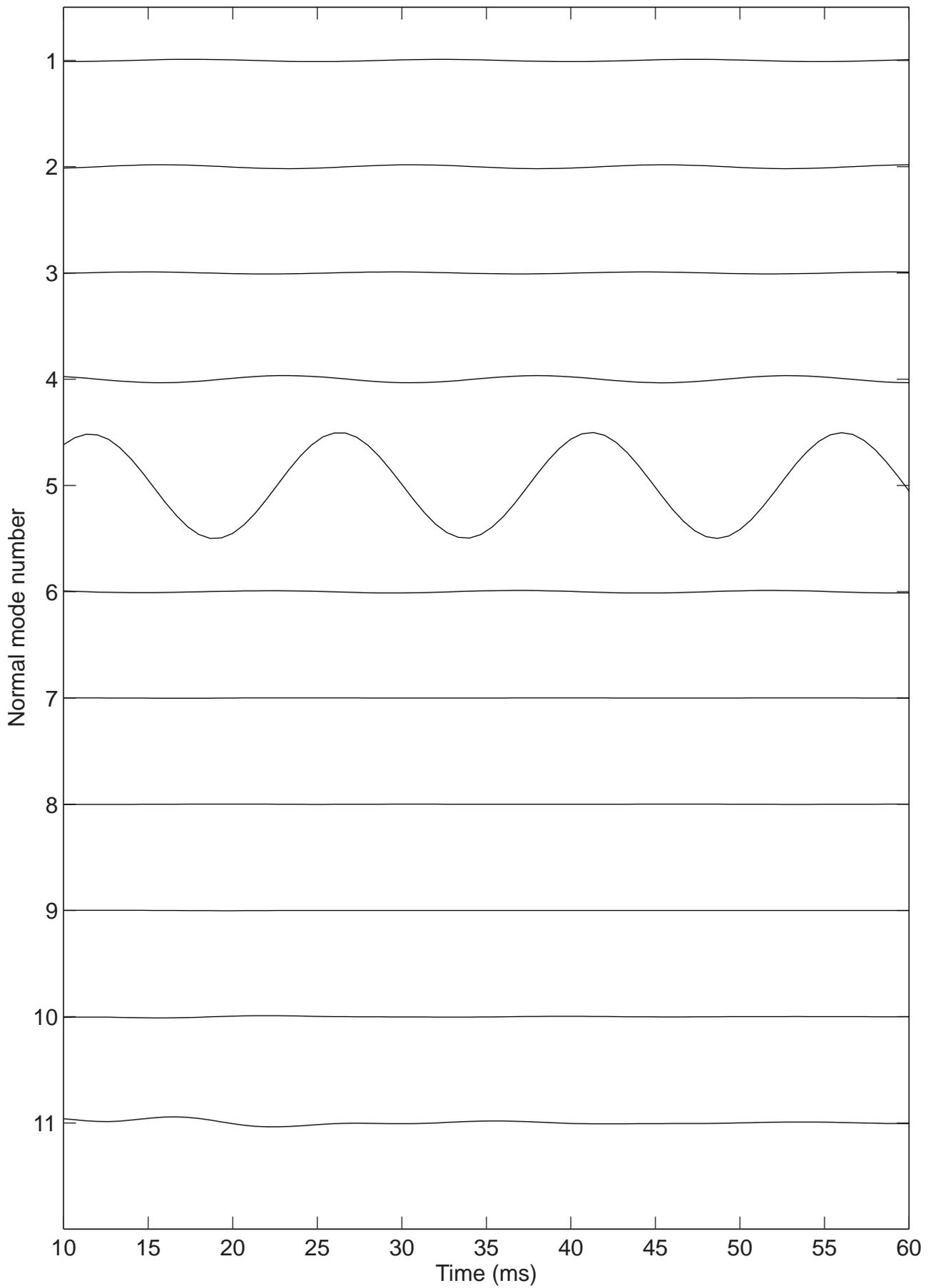
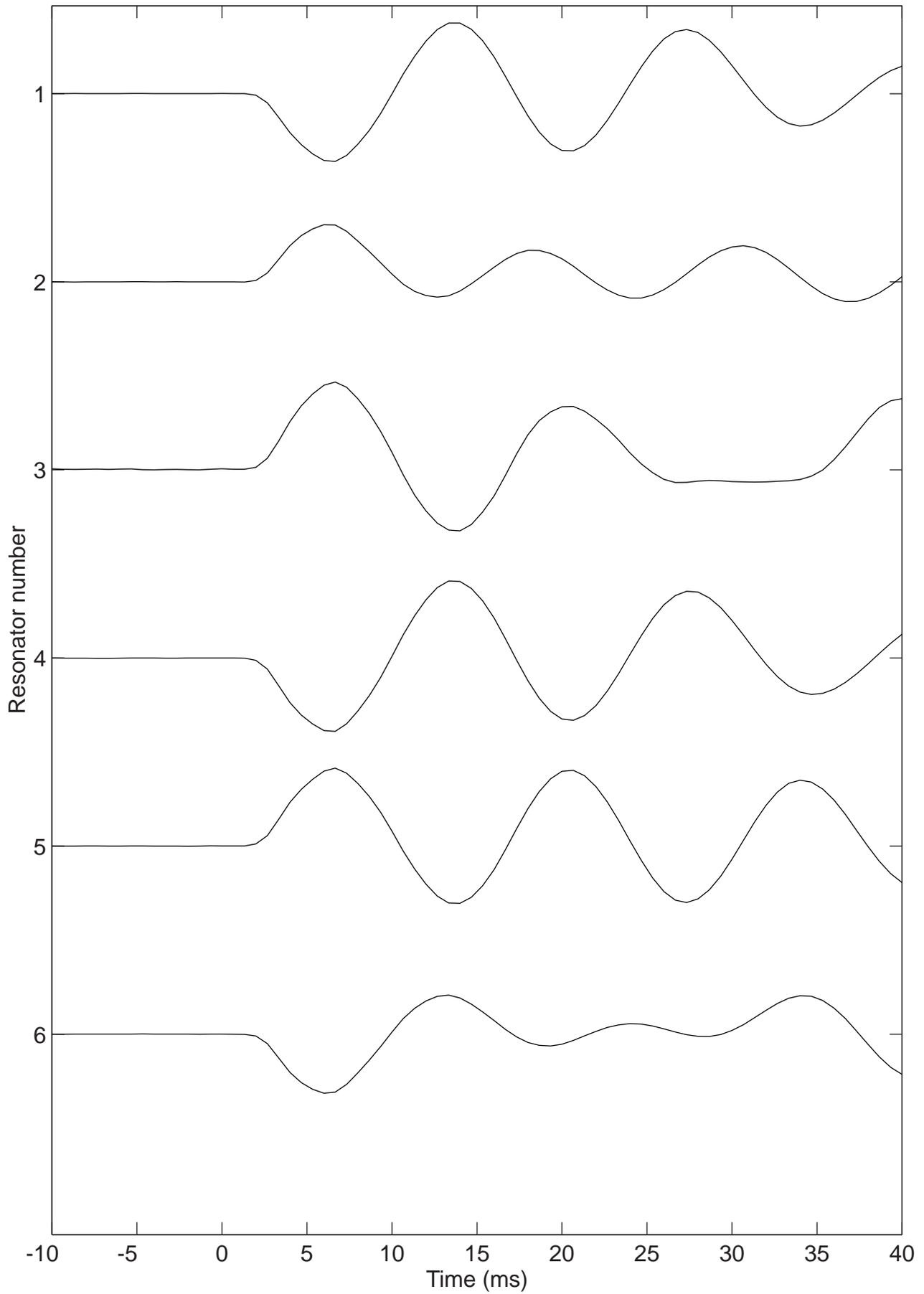
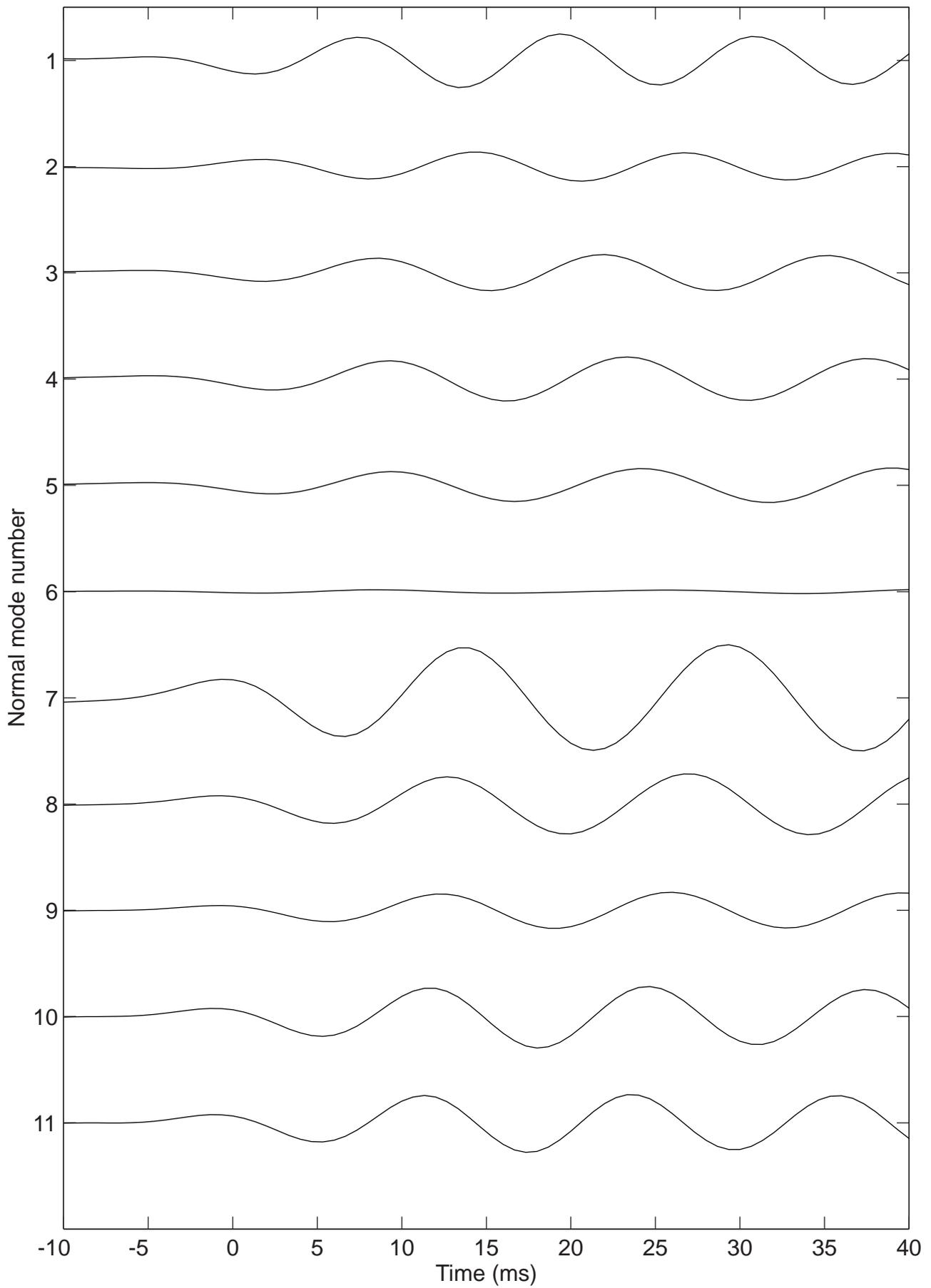
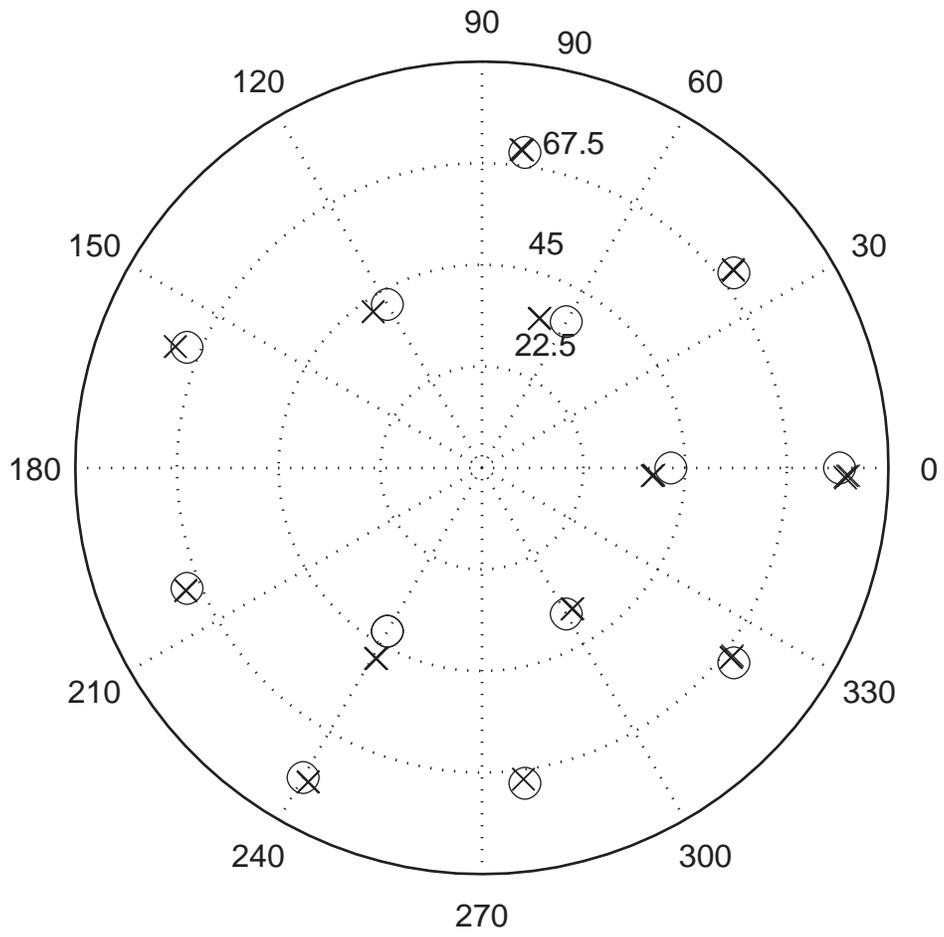
\end{document}